\documentclass[twocolumn,prd,superscriptaddress,preprintnumbers,amsmath,amssymb,nofootinbib]{revtex4}

\usepackage{amsmath}
\usepackage{amssymb}
\usepackage{slashed}
\usepackage{graphicx}
\usepackage{graphics}
\usepackage{epsfig}
\usepackage{color}
\usepackage{hyperref}
\newcommand{\Eqref}[1]{eq.~\eqref{#1}}

\newcommand{\brhoa}{\bar{\rho}_a}
\newcommand{\brhob}{\bar{\rho}_b}
\newcommand{\brhoc}{\bar{\rho}_c}
\newcommand{\blam}{\bar{\lambda}_{\phi}}
\newcommand{\patt}{\tilde{\partial}_t}

\begin{document}
\title{Quantum-gravity-induced matter self-interactions in the asymptotic-safety scenario}

\author{Astrid Eichhorn}
\affiliation{\mbox{\it Perimeter Institute for Theoretical Physics, 31 Caroline Street N, Waterloo, N2L 2Y5, Ontario, Canada}
\mbox{\it E-mail: {aeichhorn@perimeterinstitute.ca}}}

\begin{abstract} 
We investigate the high-energy properties of matter theories coupled to quantum gravity. Specifically, we show that quantum gravity fluctuations generically induce matter self-interactions in a scalar theory. 
Our calculations apply within asymptotically safe quantum gravity, where our results indicate that the UV is dominated by an interacting fixed point, with non-vanishing gravitational as well as matter couplings. In particular, momentum-dependent scalar self-interactions are non-zero and induce a non-vanishing momentum-independent scalar potential.

Furthermore we point out that terms of this type can have observable consequences in the context of scalar-field driven inflation, where they can induce potentially observable non-Gau\ss{}ianities in the CMB.

\end{abstract}

\maketitle

\section{Introduction}

In quantum gravity research, much work focuses on pure gravity, not taking into account dynamical matter degrees of freedom. However it is a priori unclear that a consistent quantum theory of gravity can straightforwardly be coupled to matter without changing some of its main properties. In other words, a consistent quantisation of gravity might require the inclusion of matter degrees of freedom from the start.
 It is thus mandatory to study the complete system of gravitational and matter degrees of freedom in a quantum theory. Then dynamical quantum gravity effects might alter some properties of the matter sector. 
Most importantly, quantum gravity fluctuations will typically generate new operators for the matter fields in the effective action. This effect occurs even if the matter theory that is coupled to gravity is just a free theory. It is the main goal of this paper to point out that to preserve asymptotic freedom in a matter theory that is coupled to quantum gravity is highly non-trivial. We will show that quantum gravity fluctuations will induce non-vanishing matter couplings, and thus not allow the matter theory to remain asymptotically free under its coupling to gravity.
We will exemplify this by considering a simple theory of a free scalar field coupled to gravity, where we point out that the so-called Gau\ss{}ian matter fixed point of asymptotically safe quantum gravity does not exist, instead all existing fixed points have non-vanishing matter self-interactions.
 Extensions of this work to the case of non-abelian gauge fields are discussed in the outlook.\newline\\
We investigate this question in a framework, in which quantum gravitational degrees of freedom are carried by the metric. Irrespective of the UV completion for gravity, such a description in terms of the metric holds within the effective-field theory framework \cite{Burgess:2003jk,Donoghue:1993eb}. It thus holds at scales, presumably below the Planck scale, where the microscopic degrees of freedom of gravity can be integrated out and traded for effective degrees of freedom carried by the metric. Thereby many UV completions for gravity can be analysed within one single framework, and observable consequences, e.g. for the CMB, can be studied.

Among the candidates for a quantum theory of gravity there is even one where the parameterisation of quantum gravitational degrees of freedom in terms of the metric in a continuum quantum field theory holds up to arbitrarily high momentum scales, namely asymptotic safety \cite{Weinberg:1980gg}. It allows to construct a predictive, continuum quantum field theory of the metric within the path-integral framework. The UV finiteness of observable quantities follows from the existence of an interacting fixed point in the running couplings, i.e. a zero in their $\beta$ functions. In the vicinity of the fixed point, the theory becomes scale-free, thus allowing for a UV limit without any divergences, in which all dimensionless couplings approach their fixed-point values. A fully non-perturbative formulation of the functional Renormalisation 
Group (FRG) \cite{Wetterich:1993yh}, for reviews see \cite{Berges:2000ew,Polonyi:2001se,Pawlowski:2005xe,Gies:2006wv,Rosten:2010vm}, has 
allowed to collect a 
substantial amount of evidence for the existence of the fixed point \cite{Reuter:1996cp,Dou:1997fg,Lauscher:2001ya,Reuter:2001ag,
Lauscher:2002sq,Litim:2003vp,Lauscher:2005qz,Codello:2006in,Fischer:2006fz,Codello:2008vh,Benedetti:2009rx,
Eichhorn:2009ah,Groh:2010ta,Eichhorn:2010tb,Manrique:2010am,Manrique:2011jc,Donkin:2012ud}, and its compatibility with Standard 
Model matter \cite{Percacci:2002ie,Percacci:2003jz, Daum:2010bc,Daum:2009dn,Harst:2011zx,Folkerts:2011jz,Eichhorn:2011pc, Eichhorn:2011ec}, even allowing for a possibility to solve the triviality 
problem in QED and the Higgs sector \cite{Harst:2011zx,Zanusso:2009bs,Vacca:2010mj,Shaposhnikov:2009pv}, for reviews see \cite{Niedermaier:2006ns,Niedermaier:2006wt,Percacci:2007sz,ASreviews, Reuter:2012id}. 
For further indications for the existence of the fixed point, see also 
\cite{Smolin:1981rm,Christensen:1978sc,Gastmans:1977ad,Niedermaier:2010zz,Hamber:2009zz,Ambjorn:2010rx}.

Such a scenario faces the following challenge: If the fixed point exists, the theory can only be predictive, if the fixed point has a finite number of UV-attractive directions (in the case of the non-interacting, Gau\ss{}ian fixed point, these are exactly given by the couplings with positive or vanishing canonical dimension). Since the fixed point is interacting, the determination of these relevant directions\footnote{The relevant directions are determined from the critical exponents $\theta_i$, which denote the negative eigenvalues of the stability matrix
\begin{equation}
\theta_i = {\rm eig} \left(-\frac{\partial \beta_{g_i}}{\partial_{g_j}} \right)\Big|_{g_{\ast}}.
\end{equation}
 In the vicinity of a fixed point, the solution to the linearised flow equation is given by
\begin{equation}
g_i (k)= g_{i\,\ast}+ \sum_n C_n V_i^n \left( \frac{k}{k_0}
\right)^{-\theta_n},
\end{equation}
with constants of integration $C_n$ and $V^n$ the $n$th eigenvector of the stability matrix. $k_0$ denotes a reference scale. (This reflects the fact that the RG flow cannot predict a numerical value for a physical scale. Thus scales such as the transition scale to the fixed-point regime have to be fixed from observations.)
Thus $\theta_i>0$ implies that towards the IR, the coupling flows away from its fixed-point value, and its IR value corresponds to a free parameter of the theory, that has to be determined by experiment.} is technically challenging. Studies within truncated RG flows without matter degrees of freedom indicate that asymptotically safe quantum gravity has at least 3 relevant couplings, and therefore more than 3 free parameters \cite{Codello:2008vh,Benedetti:2009rx,Eichhorn:2010tb,Manrique:2010am}.\\
Including matter into the theory, there are three distinct scenarios with regard to predictivity: Either, matter comes with the relevant couplings it already has in the Standard Model, i.e. without coupling it to gravity. A more exciting possibility is, that gravity might actually turn some or all of these couplings into irrelevant, and thus predictable couplings. Indications for a possible realisation of this scenario in QED are discovered in \cite{Harst:2011zx}. \\
Here we point out that gravity might actually induce further matter interactions which correspond to relevant couplings. Thus a viable quantum theory of dynamical gravitational and matter degrees of freedom might actually have \emph{more} relevant couplings, then the sum of relevant couplings in both matter and gravity theories taken separately. We will discuss how this scenario might be realised within a truncation of the full matter and gravity effective action.
\newline
Further, the fact that residual interactions exist at the fixed point will of course be crucial for physical predictions obtained from this scenario. In particular, the question if matter couplings have residual interactions in the far UV can open a window into the realm of quantum gravity phenomenology. In fact, predictions for high-energy scattering experiments can be obtained from the effective action, see, e.g. \cite{Litim:2007iu,Gerwick:2010kq,Gerwick:2011jw,Dobrich:2012nv}, and crucially depend on the type of operators that are present in the action.
Here, we use this framework to address the question how different quantum gravity proposals could be distinguished by the amount of non-Gau\ss{}ianity that they induce in the CMB, assuming an inflationary scenario with a single scalar field satisfying slow-roll conditions.

In the following we will examine the structure of the fixed point, i.e. its number of relevant directions, and its non-vanishing couplings, in the context of a scalar matter theory. We will show that a non-vanishing value for the Newton coupling induces non-vanishing matter self-interactions. This conclusion holds in particular at the fixed point in the context of the asymptotic-safety scenario, but it is also valid within an effective field-theory framework for quantum gravity. In the context of a scalar theory, asymptotic freedom is usually not expected to hold in any case; but in a broader context, our calculation exemplifies a general mechanism, that presumably also applies in the case of asymptotically free Yang-Mills theories.\newline\\
Before we present the details of the calculation, let us explain the basic idea underlying it:
The reason why a free matter theory does not remain such when coupled to gravity is simple to understand: Consider any free matter theory which is coupled to gravity minimally. Then the integration measure contains a factor $\sqrt{g}$ \footnote{Note that we work in a Euclidean setting here, however our arguments carry over to a Lorentzian setting. Explicit RG calculations indicate that results regarding asymptotically safe quantum gravity hold also in the Lorentzian case \cite{Manrique:2011jc}.}. An expansion of $\sqrt{g}$ around any background metric produces an arbitrary number of vertices.
Thus the matter theory is not a free theory any more, instead it contains interaction vertices with the gravitational field of arbitrarily high power. It is then straightforward to construct loop diagrams which contain metric loops only, and an arbitrary number of external matter legs. These loop diagrams generate further effective matter-gravity interactions, and most importantly, matter self-interactions.\\
The situation can be compared to QED, where no photon self-interactions are present at the classical, microscopic level, but fermionic loops induce photon-photon-interactions already in the one-loop effective action.

As an example, consider a simple scalar theory described by the following effective action
\begin{equation}
\Gamma_k=  \frac{1}{2} Z_{\phi} (k) \int d^4x\, \sqrt{g}g^{\mu \nu}\partial_{\mu} \phi \partial_{\nu}\phi
\end{equation}
with a momentum-scale dependent wave-function renormalisation $Z_{\phi}(k)$. The momentum scale $k$ indicates an infrared cutoff, such that all fluctuations in the path integral with momenta $p^2 > k^2$ have been integrated out.\\
Here we have made use of the fact that the covariant derivative simplifies when acting on a scalar field. 

Four-scalar interactions are then generated by the following loop diagrams:\newline
\begin{figure}[!here]
\includegraphics[scale=0.1]{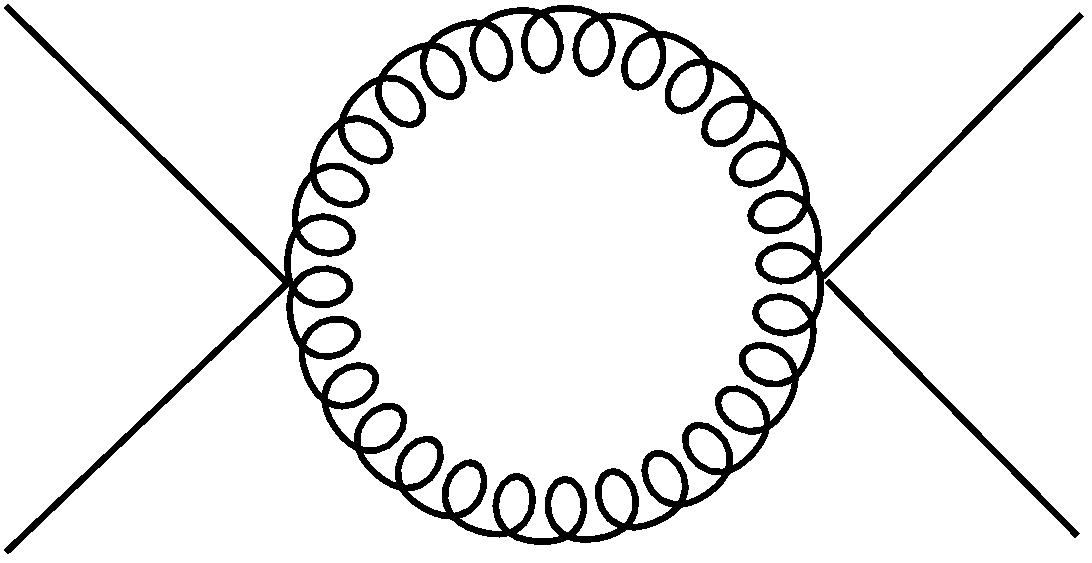}
\quad \quad
\includegraphics[scale=0.1]{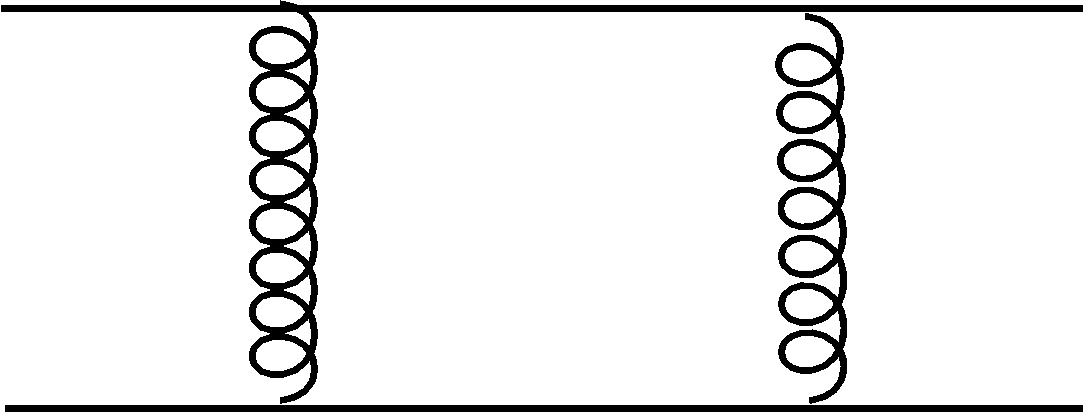} 
\caption{The scalar-gravity-interaction vertices that are generated by the kinetic term for the scalar field allow to construct these two one-loop diagrams that generate scalar self-interactions.\label{diags1}}
\end{figure}\\
Similar diagrams induce higher $\phi^{2n}$ interactions. 
 It follows from dimensional analysis that such diagrams will generate momentum-dependent scalar self-interactions: Depending on the convention one chooses where to have the scale-dependent Newton coupling $G_N(k)$ appear, either the graviton propagator will be $\sim G_N$, or each vertex will be $\sim G_N$ for the two-vertex diagram, and $\sim \sqrt{G_N}$ for the four-vertex diagram. Thus the contribution will always generate a term $\sim G^2_N$, so four powers of momenta are necessary to compensate the dimensionality of the Newton coupling. Thereby the first term that is induced in the effective action will be quartic in the scalar field and in the momentum. The associated coupling will later be called $\bar{\rho}$.
 Thus gravity fluctuations do not directly induce a running in the scalar potential, as has been observed in \cite{Narain:2009fy,Narain:2009gb}. However, as we will point out, there is an indirect effect: The metric-induced momentum-dependent scalar self interactions contribute to the running of the scalar potential, and induce a non-vanishing $\lambda_{\phi} \phi^4$ term.

\begin{figure}[!here]
\includegraphics[width=\linewidth]{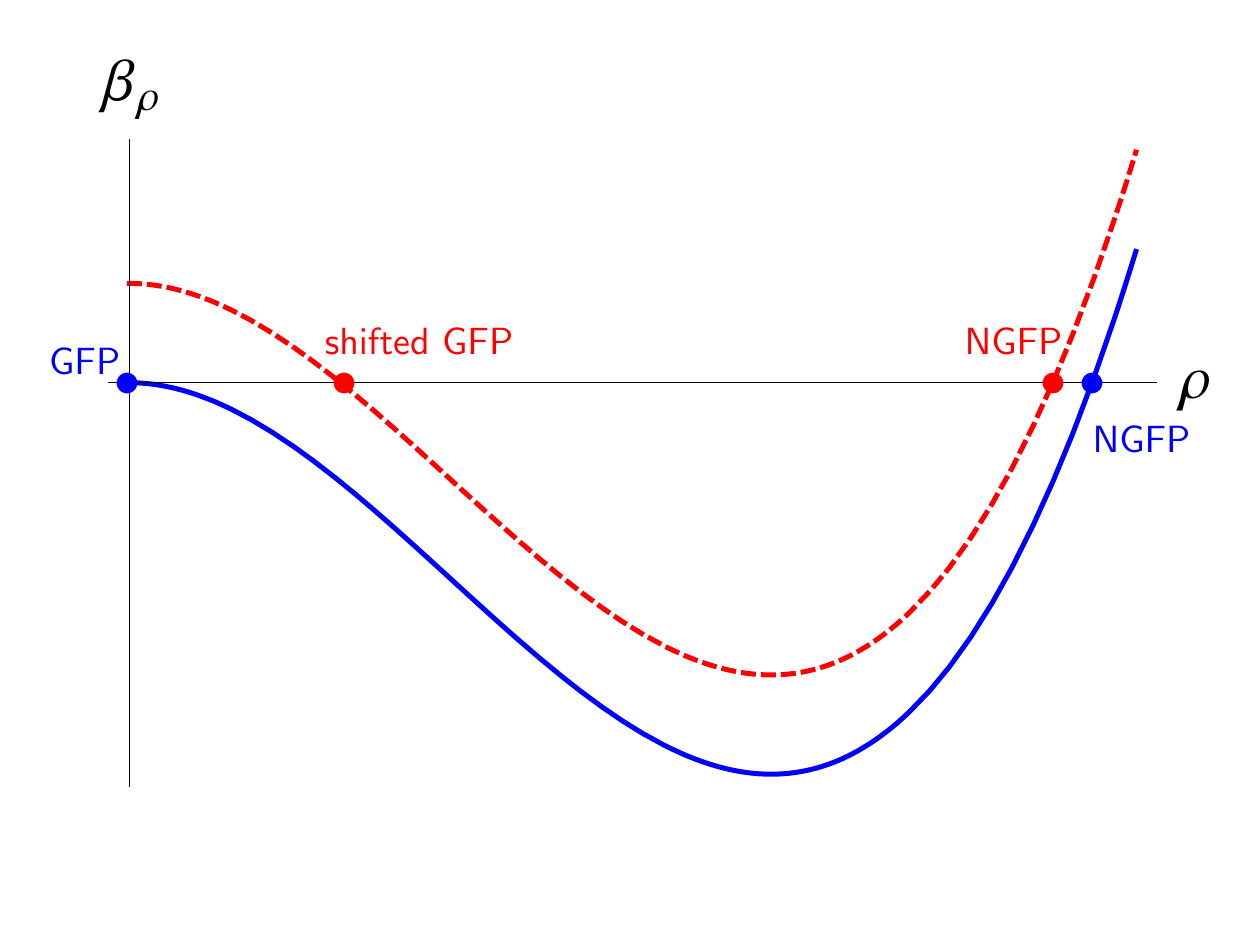}
\caption{\label{shiftedGFP} Here we exemplify a matter $\beta$ function for the coupling $\rho$: Without coupling to gravity (blue line), this $\beta$ function will admit a Gau\ss{}ian fixed point, as well as possibly a non-Gau\ss{}ian one. Including gravity fluctuations, a term $\sim G$ will be added to the $\beta$ function, thus shifting the Gau\ss{}ian to an interacting, non-Gau\ss{}ian fixed point (red dashed line). A second non-Gau\ss{}ian fixed point may or may not exist.}
\end{figure}

The crucial point about the diagrams in fig.~\ref{diags1} is, that they yield contributions $\sim G_N^{m}$ where m is the number of metric propagators, and are independent of the coupling that they generate. In the example above, we thus get the following contribution to the $\beta$ function of the dimensionful scalar coupling $\bar{\rho}$ (see \Eqref{mattertrunc}):
\begin{equation}
\beta_{\bar{\rho}} =  G_N^2 f(\lambda)+...,
\end{equation}
where $f(\lambda)$ is a function fo the cosmological constant and further terms are $\sim \bar{\rho}$, $\sim \bar{\rho}^2$ and proportional to further matter couplings. The main point is, that, even if we set all matter couplings to zero, i.e. we start with a free theory, then metric fluctuations generate interaction terms for the matter, see also fig.~\ref{shiftedGFP}.

Most importantly, metric fluctuations can add terms to the matter $\beta$ functions which lead to a vanishing of all fixed points. If such a finding where to be confirmed in untruncated theory space, the asymptotic-safety scenario would not be compatible with the existence of this particular type of matter. Within a truncation, such a definite conclusion can obviously not be drawn.

\section{Set-up of the calculation}\label{calculation}
To evaluate the $\beta$ functions of the running couplings we require a tool that is applicable in the perturbative as well as the non-perturbative regime. We also aim for an analytic continuum calculation. We employ the functional Renormalisation Group, where the Wetterich equation \cite{Wetterich:1993yh} allows to evaluate $\beta$ functions even in the non-perturbative regime. Introducing an infrared (IR) mass-like regulator function $R_k(p)$ suppresses IR modes (with $p^2 <k^2$) in the generating functional. The scale-dependent effective action $\Gamma_k$ then contains the effect of quantum fluctuations above the scale $k$ only. Its scale-dependence is given by the following functional differential equation:
\begin{equation}
\partial_t \Gamma_k= \frac{1}{2} {\rm STr} \left(\Gamma_k^{(2)}+R_k \right)^{-1}\partial_t R_k.
\end{equation}
Herein $\partial_t = k\, \partial_k$ and $\Gamma_k^{(2)}$ is matrix-valued in field space and denotes the second functional derivative of the effective action with respect to the fields. Adding the mass-like regulator and taking the inverse yields the full propagator.
The supertrace contains a trace over all indices; in the case of a continuous momentum variable it thus implies an integration over the momentum.
On the technical side, the main advantage of this equation is its one-loop form, since it can be written as the supertrace over the full propagator, with the regulator insertion  $\partial_t R_k$ in the loop. Nevertheless it is crucial to stress that it also yields higher terms in a perturbative expansion, since it depends on the full, field- and momentum-dependent propagator, and not just on the perturbative propagator, see, e.g. \cite{Papenbrock:1994kf}. 

For the gravitational part we work in the background field
 formalism \cite{Abbott:1980hw}, where the full metric is
split according to
\begin{equation}
 g_{\mu \nu}= \bar{g}_{\mu \nu}+ h_{\mu \nu},
\end{equation}
where this split does not imply that we consider only small fluctuations
around, e.g. a flat background. Within the FRG approach we have access to
physics also in the fully non-perturbative regime.  The background-field formalism, being
highly useful in non-abelian gauge theories (see, e.g. \cite{Pawlowski:2005xe,Gies:2006wv}), is mandatory in gravity, since the
background metric allows for a meaningful notion of "high-momentum" and
"low-momentum" modes as implied by the spectrum of the background covariant
Laplacian.
Within Yang-Mills theories the $\beta$ functions are independent of the choice of background, as long as it serves to distinguish the different operators in the truncation. Within gravity, an exception exists, as the topology of the background can alter the spectrum of fluctuations in the infrared. Thus only the UV behaviour of the running couplings is independent of the choice of background, whereas topologically distinct backgrounds can lead to a different IR behaviour \cite{Reuter:2008wj}. As we are only interested in the $\beta$ functions as they pertain to the UV, we can choose different backgrounds for the gravitational and the matter $\beta$ functions.

In the context of an interacting theory, the Wetterich equation is usually applied to a truncated theory space. We thus do not keep all infinitely many terms that are part of the full effective action, but concentrate on a (typically) finite subset. In our case, where we are interested in demonstrating that from a free matter theory coupled to gravity matter self-interactions will be generated, we choose a truncation of the following form:
\begin{eqnarray}
\Gamma_k &=& \Gamma_{k\,\mathrm{EH}} + \Gamma_{k\,\mathrm{gf}}+\frac{1}{2}Z_{\phi}(k) \int d^4x \sqrt{g} \,g^{\mu \nu}\partial_{\mu}\phi \partial_{\nu}\phi \nonumber\\
&+&\bar{\rho}_a(k)\int d^4x \sqrt{g} \,g^{\mu \nu}g^{\kappa \lambda}\,\partial_{\mu}\phi \partial_{\nu}\phi\, \partial_{\kappa}\phi\partial_{\lambda}\phi\nonumber\\
&+&  \bar{\rho}_b (k) \int d^4x \sqrt{g}\, \phi^2 (g^{\mu \nu}\nabla_{\mu}\partial_{\nu} \phi)\, (g^{\kappa \lambda} \nabla_{\kappa} \partial_{\lambda} \phi) \nonumber\\
&+& \bar{\rho}_c(k) \int d^4x  \sqrt{g} \,(g^{\mu \nu}\partial_{\mu} \phi \partial_{\nu}\phi)\, (\phi g^{\kappa \lambda} \nabla_{\kappa} \partial_{\lambda} \phi)\nonumber\\
&+& \bar{\lambda}_{\phi} (k)\int d^4x \sqrt{g}\, \phi^4
,\label{mattertrunc}
\end{eqnarray}
where $\nabla_{\mu}$ denotes the usual covariant derivative.
It is important to note that we have included all independent operators of fourth order in derivatives and fields. All other operators of the same order can be rewritten as a linear combination of the above ones, and possibly additional terms that depend on the curvature.

In order to demonstrate that metric fluctuations also remove the Gau\ss{}ian fixed point in the scalar potential, we include the $\bar{\lambda}_{\phi}$ coupling, which is to be distinguished from the cosmological constant $\bar{\lambda}$ by its subscript.

The Einstein-Hilbert and the gauge-fixing term read:
\begin{eqnarray}
\Gamma_{k\,\mathrm{EH}}&=& 2 \bar{\kappa}^2 Z_{\text{N}} (k)\int 
d^4 x \sqrt{g}(-R+ 2 \bar{\lambda}(k))\label{eq:GEH},\\
\Gamma_{k\,\mathrm{gf}}&=& \frac{Z_{\text{N}}(k)}{2\alpha}\int d^4 x
\sqrt{\bar g}\, \bar{g}^{\mu \nu}F_{\mu}[\bar{g}, h]F_{\nu}[\bar{g},h]\label{eq:Ggf},
\end{eqnarray}
with
\begin{equation}
 F_{\mu}[\bar{g}, h]= \sqrt{2} \bar{\kappa} \left(\bar{D}^{\nu}h_{\mu
   \nu}-\frac{1+\rho_{\rm gh}}{4}\bar{D}_{\mu}h^{\nu}{}_{\nu} \right). 
\end{equation}
Herein, $\bar{\kappa}= (32 \pi G_{\text{N}})^{-\frac{1}{2}}$ is related to the
bare Newton constant $G_{\text{N}}$ and $\bar\lambda$ is the cosmological constant.
The gauge-fixing term is supplemented by an appropriate ghost term. As within this truncation ghosts do not couple to the scalar field\footnote{Note however that in general the existence of such coupling should be expected, as it is induced by diagrams similar to those in fig.~\ref{diags1}.}, we can neglect the ghost term in the calculation of the scalar $\beta$ functions. We work in Landau deWitt gauge $\rho_{\rm gh} \rightarrow \alpha$, $\alpha \rightarrow 0$ here. In the following, we employ a decomposition of the metric fluctuations into irreducible components under the Poincare group. In our choice of gauge, only the transverse traceless mode $h_{\mu \nu}^{TT}$ (with $\bar{D}^{\mu}h_{\mu \nu}^{TT}=0$ and $\bar{g}^{\mu \nu} h_{\mu \nu}^{TT}=0$) and the trace mode $h = \bar{g}^{\mu \nu}h_{\mu \nu}$ can contribute to the running in the matter sector.

\begin{figure}[!here]
\includegraphics[width=\linewidth]{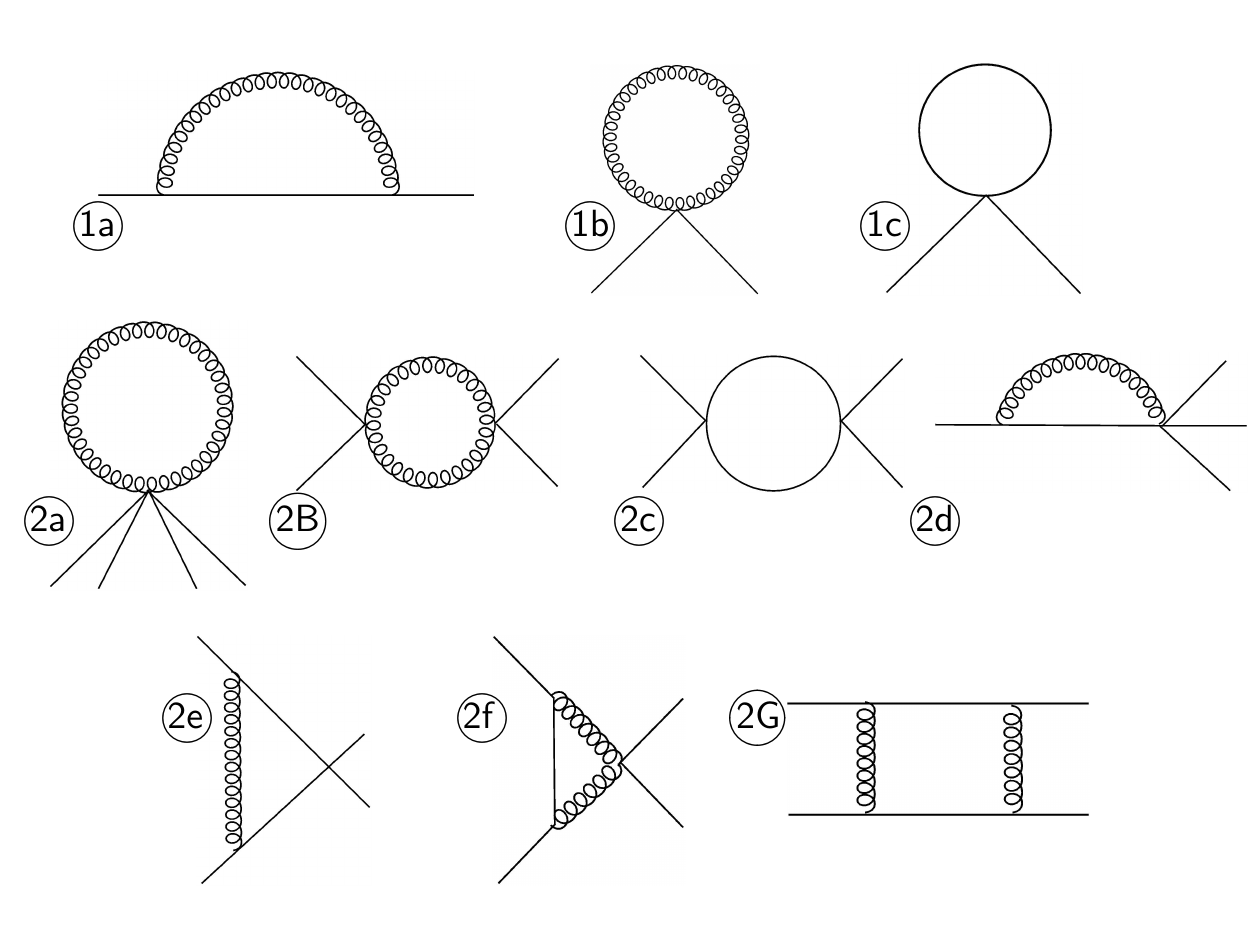}
\caption{\label{alldiags}Here we show the diagrammatric expansion of the Wetterich equation, projected onto the $\phi^2$ and $\phi^4$ subspace of theory space, respectively. Diagrams 1a, 1b and 1c contribute to $\eta_{\phi}= - \partial_t \ln Z_{\phi}$, whereas all diagrams 2 in principle contribute to $\beta_{\rho_i}$ and $\beta_{\lambda}$.
Each loop contains a regulator insertion $\partial_t R_k$ that can be found on any of the internal lines.}
\end{figure}

To evaluate matter $\beta$ functions, it is useful to proceed as follows:
Splitting $\Gamma_k^{(2)}+R_k =\mathcal{P}_k+\mathcal{F}_k$, where all scalar-field dependent
terms enter the fluctuation matrix $\mathcal{F}_k$, we may
now expand the right-hand side of the flow equation as
follows:
\begin{eqnarray}
 \partial_t \Gamma_k&=& \frac{1}{2}{\rm STr} \{
 [\Gamma_k^{(2)}+R_k]^{-1}(\partial_t R_k)\}\label{eq:flowexp}\\
&=& \frac{1}{2} {\rm STr}\, \tilde{\partial}_t\ln
\mathcal{P}_k
+\frac{1}{2}\sum_{n=1}^{\infty}\frac{(-1)^{n-1}}{n} {\rm
  STr}\,
\tilde{\partial}_t(\mathcal{P}_k^{-1}\mathcal{F}_k)^n,
\nonumber
\end{eqnarray}
where the derivative $\tilde{\partial}_t$ in the second line by definition
acts only on the $k$ dependence of the regulator, $\tilde{\partial}_t=\int \partial_t R_k\frac{\delta}{\delta R_k}$, see appendix \ref{partialtreg}. Since each
factor of $\mathcal {F}_k$ contains a
coupling to external fields, this expansion simply corresponds to an expansion
in the number of vertices. Thus we can straightforwardly write down the diagrammatic expansion of a $\beta$ function, see fig.~\ref{alldiags}. Herein, the internal propagator lines contain the mass-like regulator-function, and in the case of the graviton, also depend on the cosmological constant. In order to project onto the anomalous dimension $\eta_{\phi} = - \partial_t \ln Z_{\phi}(k)$, we can terminate the expansion at second order in the vertices, since all further terms in the expansion must have more than two external scalar fields. Similarly, the evaluation of the $\partial_t \bar{\rho}_i(k)$  for $i= a,b,c$ and $\partial_t \bar{\lambda}_{\phi}$ requires terms up to $\left(\mathcal{P}^{-1}\mathcal{F}\right)^4$, see fig.~\ref{alldiags}.

Clearly a flat background, $\bar{g}_{\mu \nu}= \delta_{\mu \nu}$ is fully sufficient to project onto the matter $\beta$ functions, and thus preferable for reasons of technical simplicity.

By using a regulator of the form 
\begin{equation}
R_k = \Gamma_k^{(2)}\Big|_{\phi=0}r\left(\frac{ \Gamma_k^{(2)}\Big|_{\phi=0}(p^2)}{k^2}\right),\label{regulator}
\end{equation}
it is then straightforward to find which type of contribution to the $\beta$ function a particular diagram corresponds to: Each graviton propagator comes with a factor of $G_N(k)$. Thus, e.g. the two-vertex contribution 2B, see fig.~\ref{diags1} yields a contribution $\sim Z_{\phi}(k)^2\, G_N^2$, since the vertex is $\sim Z_{\phi}(k)$, and each propagator $\sim G_N(k)$.

Next we introduce the anomalous dimension
\begin{equation}
\eta_{\phi}= - \partial_t \ln Z_{\phi},
\end{equation}
as well as
dimensionless and renormalised couplings
\begin{eqnarray}
\rho_i (k)&=& \bar{\rho}_i(k) \frac{k^4}{Z_{\phi}^2(k)}\,  \quad \quad i = a,b,c,\nonumber\\
\lambda_{\phi}(k)&=& \bar{\lambda}_{\phi}\frac{1}{Z_{\phi}(k)^2},\nonumber\\
g(k)&=&\frac{G_{\text{N}} k^2}{Z_{\text{N}}(k)}=  \frac{k^2}{32\pi\bar{\kappa}^2\,  Z_{\text{N}}},\nonumber\\
\lambda(k)&=& \frac{\bar{\lambda}}{k^2}.\label{dimless}
\end{eqnarray}
It is then straightforward to see that the $\beta_{\rho_i}$  have the following form
\begin{eqnarray}
&{}&\partial_t \rho_i =\beta_{\rho_i}\nonumber\\
&=& 4 \rho_i + 2 \eta_{\phi} \rho_i + c_1\, g^2 \, f_1(\lambda)+ \sum_{j= a,b,c} c_2\, g \, \rho_j \, f_2 (\lambda) \nonumber\\
&+& \!\!\!\!\sum_{j,k= a,b,c} c_3\, \rho_j \rho_k, + c_4 \, g\, \lambda_{\phi} \, f_3(\lambda) + \sum_{j=a,b,c} c_5\, \rho_j \lambda_{\phi},\label{betarhoschem}
\end{eqnarray}
where the first term arises from the canonical dimensionality. The constants $c_i$ are regulator and gauge dependent, as are the functions $f_i(\lambda)$. These quantities remain to be determined. Crucially, we see that depending on the $c_i$ and the behaviour of $f_i(\lambda)$, the point $\rho_i =0$ may \emph{not} be a fixed point of the RG flow, whereas it clearly is in the case $g=0$.\\

In contrast, the $\beta$ function for $\lambda_{\phi}$ will not have a contribution $\sim g^2$, since the diagrams 2B and 2G in fig.~\ref{alldiags} vanish if projected onto vanishing external momenta.
We thus have that
\begin{eqnarray}
\beta_{\lambda_{\phi}}&=&  2 \eta_{\phi} \lambda_{\phi} + \sum_{j= a,b,c} d_1\, g \, \rho_j \, h_1 (\lambda) + \sum_{j,k= a,b,c} d_2\, \rho_j \rho_k, \nonumber\\
&+& d_3 \, g\, \lambda_{\phi} \, h_2(\lambda) + \sum_{j=a,b,c} d_4\, \rho_j \lambda_{\phi},\label{betalamschem}
\end{eqnarray}
where $h_i(\lambda)$ are scheme-dependent functions of $\lambda$ and $d_i$ are scheme-dependent constants.
Here, the contributions $\sim \rho_i^2$ play an important role, since for $d_2 \neq 0$, these imply that $\beta_{\lambda_{\phi}} \neq 0$ at $\lambda_{\phi}=0.$

Having said that the $c_i, d_i$ and $f_i(\lambda), h_i(\lambda)$ have a gauge- as well as regulator-dependence, let us note that this will induce quantitative variations in fixed-point values and critical exponents, however the existence of fixed points and the relevance of these couplings  are universal properties.

For the gravitational $\beta$ functions, it is crucial to note that there cannot be a direct contribution $\sim \rho_i, \lambda_{\phi}$, since $\Gamma_k^{(2)}$ evaluated at vanishing $\phi_i$ (in order to project onto the terms $\int d^4x \, \sqrt{g}$ and $\int d^4x\, \sqrt{g}\, R$) does not contain a term $\sim \rho_i$ or $\sim \lambda_{\phi}$. Thus the matter contribution to $\partial_t g$ and $\partial_t \lambda$ arises from the kinetic matter term only. Since $\eta_{\phi}= \eta_{\phi}(\rho_a, \rho_b,\rho_c)$, we do however get a backcoupling of the $\rho_i$ into the gravitational $\beta$ functions in this indirect way.
Here we can use that a maximally symmetric background with positive curvature suffices to evaluate these $\beta$ functions. For the metric and ghost contributions, we proceed as in \cite{Eichhorn:2009ah,Eichhorn:2010tb}. For the matter contribution, we use that for an exponential shape function it is straightforward to evaluate the functional traces via an inverse Laplace transform; for details, see appendix \ref{matterbetacontr}.

In the following, we will use a spectrally and RG adjusted regulator \cite{Gies:2002af,Litim:2002xm} of the form \eqref{regulator}
with exponential shape function $r(y)= \left( e^{y}-1\right)^{-1}$. Therein $p^2$ is understood to be a placeholder for the eigenvalues of the differential operator constituting the kinetic term, evaluated on the appropriate background. As an example, for the evaluation of $\beta_{\rho_i}$, $p^2$ is to be understood as the momentum, whereas $p^2 = - \bar{D}^2$ with $\bar{D}^2$ being the Laplacian on a maximally symmetric space of positive curvature for the evaluation of $\beta_{g}$ and $\beta_{\lambda}$. Note that our choice of regulator implies that the right-hand side of the flow equation will depend on $\eta_{\phi}$, $\eta_N$ and $\partial_t \lambda$, but not on $\partial_t \rho_i$ and $\partial_t \lambda_{\phi}$.

\begin{widetext}
To obtain the anomalous dimension, we apply the following projection rule:
\begin{eqnarray}
\eta_{\phi}= -\frac{1}{8}\left(Z_{\phi} \right)^{-1}\Bigl( \frac{\partial}{\partial p_{1\mu}}\frac{\partial}{\partial p_1^{\mu}} \int \frac{d^4 p_2}{(2 \pi)^4} \frac{\delta}{\delta \phi(p_1)} \frac{\delta}{\delta \phi(p_2)} \partial_t \Gamma_k \Bigr)\Big|_{\phi= 0, p_1=0}.
\end{eqnarray}

To evaluate the $\beta$ functions for the matter couplings we employ the following projection rules, where the numerical coefficients arise due to the differing tensor structure of the four matter couplings:
\begin{eqnarray}
\partial_t \bar{\rho}_a &=&  \left(\left(\frac{1}{384} \partial_{l_{1\mu}}   \partial_{l_{1\mu}}   \partial_{l_{2\nu}}   \partial_{l_{2\nu}} -\frac{11}{1152}\partial_{l_{1 \mu}} \partial_{l_{1\mu}}\partial_{l_{2\nu}}\partial_{l_{3\nu}}+\frac{1}{144}\partial_{l_{1\mu}}\partial_{l_{2\mu}}\partial_{l_{1\nu}}\partial_{l_{3\nu}}  \right)\int \frac{d^4 l_4}{(2 \pi)^4}\frac{\delta^4}{\delta \phi(l_1) \delta \phi(l_2)\delta \phi(l_3)\delta \phi(l_4)} \Gamma_k\right)\Big|_{\phi=0, l_i =0},\nonumber\\
\partial_t \bar{\rho}_b &=&  \left(\left(\frac{1}{384} \partial_{l_{1\mu}}   \partial_{l_{1\mu}}   \partial_{l_{2\nu}}   \partial_{l_{2\nu}} -\frac{1}{192}\partial_{l_{1 \mu}} \partial_{l_{1\mu}}\partial_{l_{2\nu}}\partial_{l_{3\nu}}+\frac{1}{192}\partial_{l_{1\mu}}\partial_{l_{2\mu}}\partial_{l_{1\nu}}\partial_{l_{3\nu}}  \right)\int \frac{d^4 l_4}{(2 \pi)^4}\frac{\delta^4}{\delta \phi(l_1) \delta \phi(l_2)\delta \phi(l_3)\delta \phi(l_4)}  \Gamma_k\right)\Big|_{\phi=0, l_i =0},\nonumber\\
\partial_t \bar{\rho}_c &=& \left(\left(\frac{1}{192} \partial_{l_{1\mu}}   \partial_{l_{1\mu}}   \partial_{l_{2\nu}}   \partial_{l_{2\nu}} -\frac{1}{48}\partial_{l_{1 \mu}} \partial_{l_{1\mu}}\partial_{l_{2\nu}}\partial_{l_{3\nu}}+\frac{1}{48}\partial_{l_{1\mu}}\partial_{l_{2\mu}}\partial_{l_{1\nu}}\partial_{l_{3\nu}}  \right)\int \frac{d^4 l_4}{(2 \pi)^4}\frac{\delta^4}{\delta \phi(l_1) \delta \phi(l_2)\delta \phi(l_3)\delta \phi(l_4)}  \Gamma_k\right)\Big|_{\phi=0, l_i =0},\nonumber\\
\partial_t \bar{\lambda}_{\phi}&=&  \frac{1}{\Omega}\frac{1}{24}\int \frac{d^4 l_4}{(2 \pi)^4}\left(\frac{\delta^4}{\delta \phi(l_1) \delta \phi(l_2)\delta \phi(l_3)\delta \phi(l_4)} \Gamma_k\right)\Big|_{\phi=0, l_i =0}.
\end{eqnarray}

\end{widetext}
For further details, and the appropriate vertices for the evaluation of the $\beta$ functions, see appendix \ref{appendix1}.
\section{Results}
\subsection{Matter sector without gravity}
Let us first set $g=0$ and thereby switch off metric fluctuations to briefly examine the pure matter truncation. As expected, it admits a Gau\ss{}ian fixed point with critical exponents $0,-4,-4,-4$ as corresponding to the canonical dimensionality. Further, we also find several non-Gau\ss{}ian fixed points in the system. These have in common, that at all of these, at least one of the couplings $\rho_i, \lambda_{\phi}$ has a negative value, thus corresponding to an unstable direction of the Euclidean action. This does however not imply that these fixed points should be discarded, since clearly higher-order operators beyond our truncation, such as $\phi^6$ operators will be induced. Thus, the Euclidean action in an extended truncation could be bounded from below even at negative values of the $\rho_i$ and $\lambda_{\phi}$. 

In tab.~\ref{fptablewograv}, we list the fixed points in this truncation; discarding further spurious fixed points since their rather large anomalous dimension suggests that terms beyond our truncation will crucially alter these.

\begin{widetext}
\begin{table}[!here]
\begin{tabular}{c|c|c|c|c|c|c|c|c}
 $\rho_{a\, \ast}$& $\rho_{b\, \ast}$& $\rho_{c\, \ast}$& $\lambda_{\phi\, \ast}$& $\eta_{\phi\, \ast}$& $\theta_{1}$& $\theta_{2}$& $\theta_{3}$& $\theta_{4}$\\
\hline
0&0&0&0&0&0&-4&-4&-4\\
-4.67&0&0&0&-0.82&4.59&1.64&0&-2.36\\
65.31 & 40.97 & 118.54&-155.51&2.87&-4.02-24.88 $i$ & -4.02+24.88 $i$& -16.47 -0.79 $i$&-16.47 +0.79 $i$
\end{tabular}
\caption{\label{fptablewograv} Here we list the fixed-point values, as well as the value of the anomalous dimension $\eta_{\phi}$ and the critical exponents at the various fixed points at $g=0$. We find the Gau\ss{}ian fixed point, as well as two non-Gau\ss{}ian ones, where the first has two relevant directions.}
\end{table}
\end{widetext}

Thus, the pure matter system admits several fixed points to construct a possible UV completion. Note that the critical exponents at the non-Gau\ss{}ian fixed points deviate significantly from the canonical dimensionality of the couplings. To determine the relevance of a coupling, only the real part of the critical exponent matters. The imaginary parts indicate that the flow approaches the fixed point in a spiral-type shape.

In particular, the second of the three fixed points even shows two relevant directions, see tab.~\ref{fptablewograv}. Hence, using this fixed point as a UV completion for the scalar theory in this truncation implies a lower level of predictivity than expected from a simple dimensional analysis.

In the following, we will discuss the fate of these fixed points under the coupling to gravity. We first focus on the Gau\ss{}ian fixed point, which becomes interacting for $g\neq0$.

\subsection{Fixed point analysis in the matter sector: shifted Gau\ss{}ian fixed point}

Our first result is that the contribution from metric fluctuations that is independent of the matter couplings $\bar{\rho}_{a,b,c}$ and $\bar{\lambda}_{\phi}$, is indeed non-zero. For unspecified regulator shape function $r(p^2/k^2)$, the contribution takes the following form:
\begin{widetext}
\begin{eqnarray}
\beta_{\bar{\rho}_a}\Big|_{\rho_{a,b,c},\lambda_{\phi}=0} 
&=&\!\frac{575}{1728} Z_{\phi}^2\int \frac{d^4p}{(2 \pi)^4}\tilde{\partial}_t \frac{1}{\left(\Gamma_{k\, h^{TT}h^{TT}}^{(2)}\left(1+r \left( \frac{\Gamma_{k\, h^{TT}h^{TT}}^{(2)}(p^2)}{k^2} \right) \right)\Big|_{\phi=0}\right)^2 }\nonumber\\
&{}&+ \frac{1}{1024} Z_{\phi}^4\int \frac{d^4p}{(2 \pi)^4}\tilde{\partial}_t \frac{1}{\left(\Gamma_{k\, hh}^{(2)}\left(1+r \left( \frac{\Gamma_{k\, hh}^{(2)}(p^2)}{k^2} \right) \right)\Big|_{\phi=0}\right)^2 }\frac{1}{\left(\Gamma_{k\, \phi \phi}^{(2)}\left(1+r \left( \frac{\Gamma_{k\, \phi \phi}^{(2)}(p^2)}{k^2} \right) \right)\Big|_{\phi=0}\right)^2 }\nonumber\\
\beta_{\bar{\rho}_b}\Big|_{\rho_{a,b,c},\lambda_{\phi}=0} 
&=&\!\frac{85}{576} Z_{\phi}^2\int \frac{d^4p}{(2 \pi)^4}\tilde{\partial}_t \frac{1}{\left(\Gamma_{k\, h^{TT}h^{TT}}^{(2)}\left(1+r \left( \frac{\Gamma_{k\, h^{TT}h^{TT}}^{(2)}(p^2)}{k^2} \right) \right)\Big|_{\phi=0}\right)^2 }\nonumber\\
\beta_{\bar{\rho}_c}\Big|_{\rho_{a,b,c},\lambda_{\phi}=0} 
&=&\!\frac{55}{96} Z_{\phi}^2\int \frac{d^4p}{(2 \pi)^4}\tilde{\partial}_t \frac{1}{\left(\Gamma_{k\, h^{TT}h^{TT}}^{(2)}\left(1+r \left( \frac{\Gamma_{k\, h^{TT}h^{TT}}^{(2)}(p^2)}{k^2} \right) \right)\Big|_{\phi=0}\right)^2}
\label{inducing_matter}
\end{eqnarray}
\end{widetext}
Here, $\Gamma_{k\, h^{TT}h^{TT}}^{(2)}$ is the scalar part of the inverse transverse traceless propagator and similarly, $\Gamma_{k\, hh}^{(2)}$ is the inverse propagator of the trace mode $h$.
Herein, all $\beta_{\bar{\rho}_i}$ receive a non-vanishing contribution from the two-vertex diagram 2B, see fig.~\ref{alldiags}, whereas only $\beta_{\bar{\rho}_a}$ receives a contribution from the four-vertex diagram 2G. Due to the different tensor structures of the operators assciated to the couplings $\bar{\rho}_i$, the projection of this diagram onto $\bar{\rho}_b$ and $\bar{\rho}_c$ vanishes.

 The two-vertex diagram only has a contribution from the transverse traceless graviton mode, due to the form of the vertices, see app.~\ref{appendix1}. Furthermore, the four-vertex diagram with internal transverse traceless gravitons has a vanishing projection onto the operators under consideration here, with four external momenta. In fact, this diagram contributes to the running of couplings of higher-order operators, i.e. operators with four scalar fields and more than four external momenta.

The effect of these contributions is simple: At vanishing matter couplings, the contributions \Eqref{inducing_matter} remain non-zero. Therefore, the right-hand side of the $\beta$ functions of the couplings $\rho_{i}$ is not zero at vanishing matter coupling.

Accordingly there cannot be a Gau\ss{}ian fixed point in the three momentum-dependent couplings $\rho_{a,b,c}$. We conclude that as soon as a free scalar theory is coupled to gravity, non-vanishing matter self-interactions are induced. 

\begin{figure}
\includegraphics[width=\linewidth]{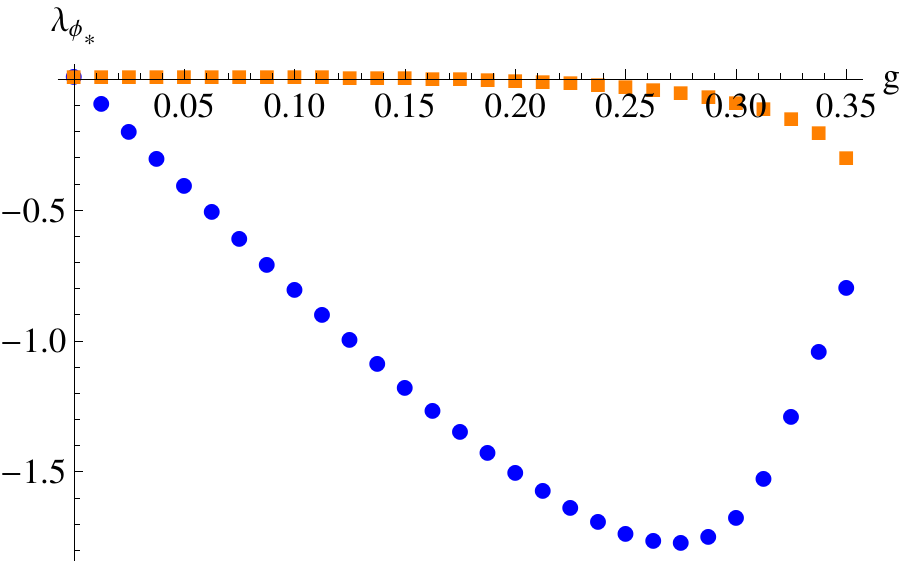}
\includegraphics[width=\linewidth]{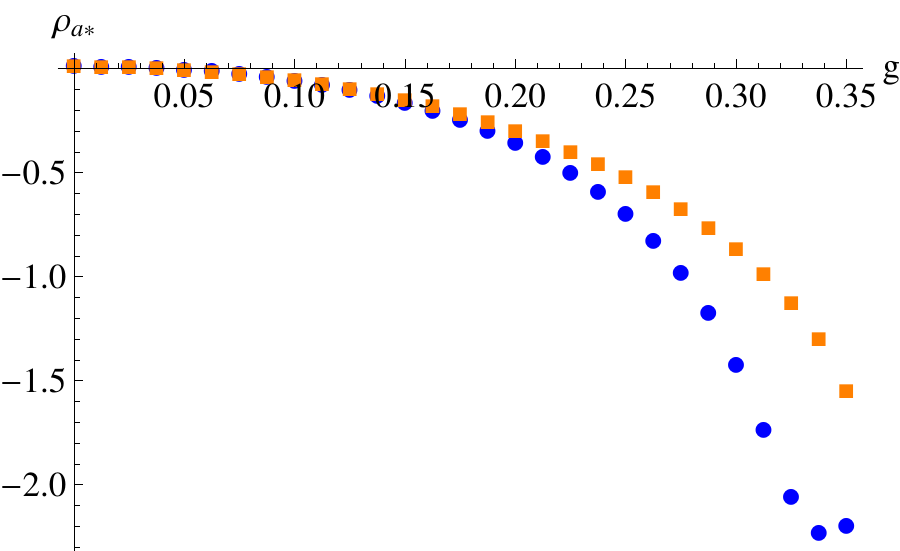}
\includegraphics[width=\linewidth]{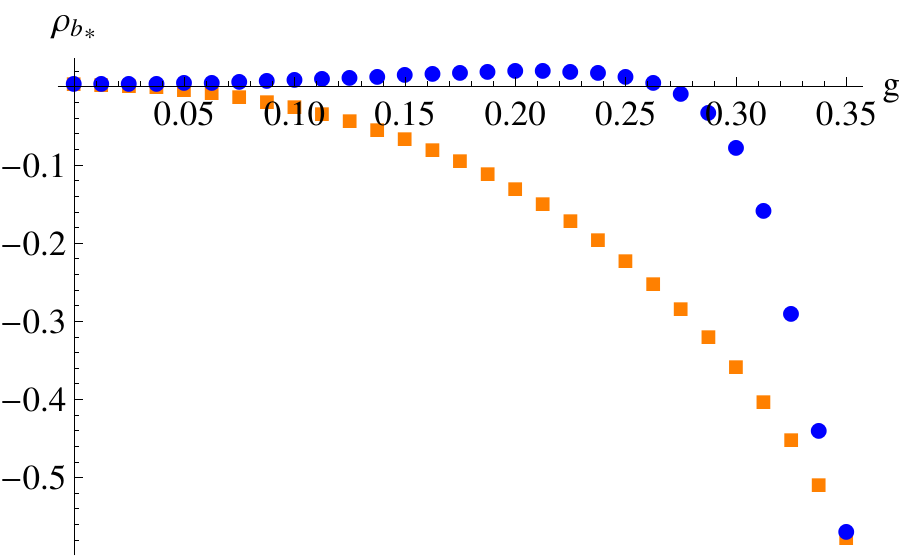}
\includegraphics[width=\linewidth]{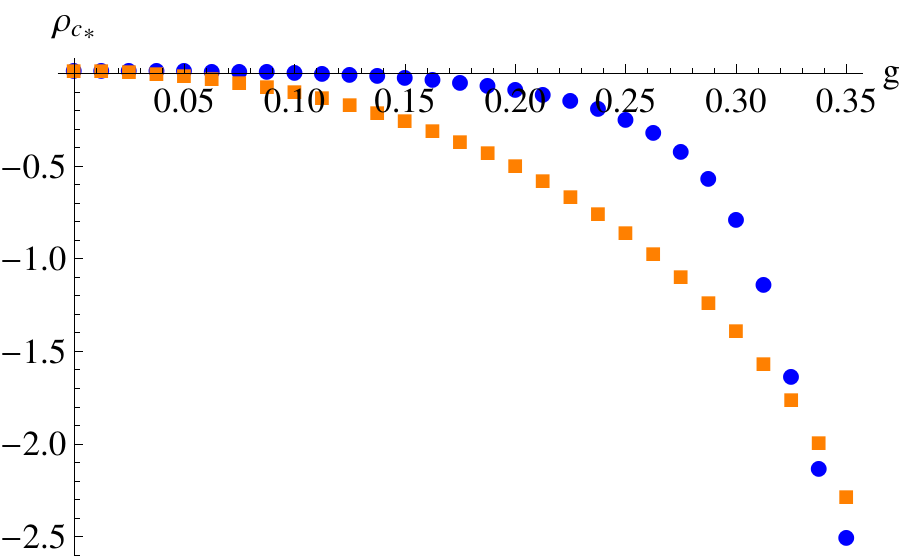}
\caption{\label{induced_FP} Here we plot the fixed-point values of the matter self-couplings as a function of $g$ for a value of the cosmological constant $\lambda = 1/10$. The blue circles show the first fixed point, the orange squares the second fixed point. The fixed point values approach each other for increasing values of $g$. Note that in fact, the values of $\lambda_{\phi}$ at the second fixed point are small, but non-zero for $g>0$.
Beyond $g =0.35$, we do not find any fixed points, i.e. the two fixed points annihilate.}
\end{figure}

This has a crucial effect on the $\beta$ function for the momentum-independent coupling $\bar{\lambda}_{\phi}$: Whereas metric fluctuations do not directly induce this term, there is a contribution arising from the purely scalar diagram 2c in fig.~\ref{alldiags}, which is $\sim \rho_b$:
\begin{eqnarray}
&{}&\beta_{\bar{\lambda}_{\phi}}\Big|_{\lambda_{\phi}=0}\\
&=&4 \bar{\rho}_b^2\int \frac{d^4p}{(2 \pi)^4}  \tilde{\partial}_t \frac{1}{\left(\Gamma_{k\, \phi \phi}^{(2)}\left(1+r \left( \frac{\Gamma_{k\,\phi \phi}^{(2)}(p^2)}{k^2} \right) \right)\Big|_{\phi=0}\right)^2 }\nonumber
\end{eqnarray}
Herein, $\Gamma_{k\, \phi \phi}^{(2)}$ is the inverse scalar propagator.
This contribution arises, as the contribution $\sim \rho_b$ to the four-scalar vertex clearly contains a part which is proportional to the momenta of two of the four scalar fields only, see app.~\ref{appendix1}. Constructing the diagram 2c in fig.~\ref{alldiags}, there will accordingly be a non-vanishing contribution if the momenta of the external fields are taken to zero. This is precisely the non-vanishing contribution to $\beta_{\lambda_{\phi}}$. We thus observe that at $\lambda_{\phi}=0$, the right-hand side of $\beta_{\lambda_{\phi}}$ is non-zero, and thus this $\beta$ function does not admit a Gau\ss{}ian fixed point, see fig.~\ref{induced_FP}.

To summarise, metric fluctuations yield a non-vanishing contribution to momentum-dependent self-interactions parameterised by the couplings $\rho_{a,b,c}$. Depending on the tensor structure, these can actually yield contributions to $\beta_{\lambda_{\phi}}$. This implies that $\lambda_{\phi}=0$ will not be a fixed point of the matter $\beta$ functions in the presence of metric fluctuations, since then $\rho_b=0$ is \emph{not} a fixed point and accordingly $\beta_{\lambda} \neq 0 $ at $\lambda_{\phi} =0$.

Thus our results imply that metric fluctuations induce a non-vanishing potential for scalar matter, since momentum-dependent as well as momentum-independent couplings will be induced.

Most importantly, this means that canonical power-counting arguments in the matter sector will not hold: If the system of $\beta$ functions admits a fixed point, it can only be a non-Gau\ss{}ian one. Therefore the critical exponents will not be given by the canonical dimensionality of the couplings, but will receive additional contributions from quantum fluctuations and have non-vanishing anomalous dimensions. This is particularly interesting in the case of power-counting marginal couplings such as $\lambda_{\phi}$, which will be turned into a relevant or irrelevant coupling.

In the following, we will illustrate this, by looking at the full system of $\beta$ functions for $\rho_{a,b,c}$ and $\lambda_{\phi}$. Here, we treat $g$ and $\lambda$ as parameters, and set $\eta_{N}=-2$ and $\partial_t \lambda =0$.

We investigate the behaviour of the Gau\ss{}ian fixed point as a function of $g$. For increasing $g$ and $\lambda$ it will clearly be shifted. Indeed we find two fixed points which are connected to the Gau\ss{}ian one continuously as a function of $g$; for the examination of further non-Gau\ss{}ian fixed points, see section \ref{fulltruncation}.

As expected, at $g=0$, where metric fluctuations are switched off, the system $\beta_{\rho_{a,b,c}}, \beta_{\lambda_{\phi}}$ admits a Gau\ss{}ian fixed point with critical exponents $\theta_{a,b,c}=-4$ and $\theta_{\lambda_{\phi}}=0$, corresponding precisely to the power-counting dimensionality of these couplings.

\begin{figure}
\includegraphics[width=\linewidth]{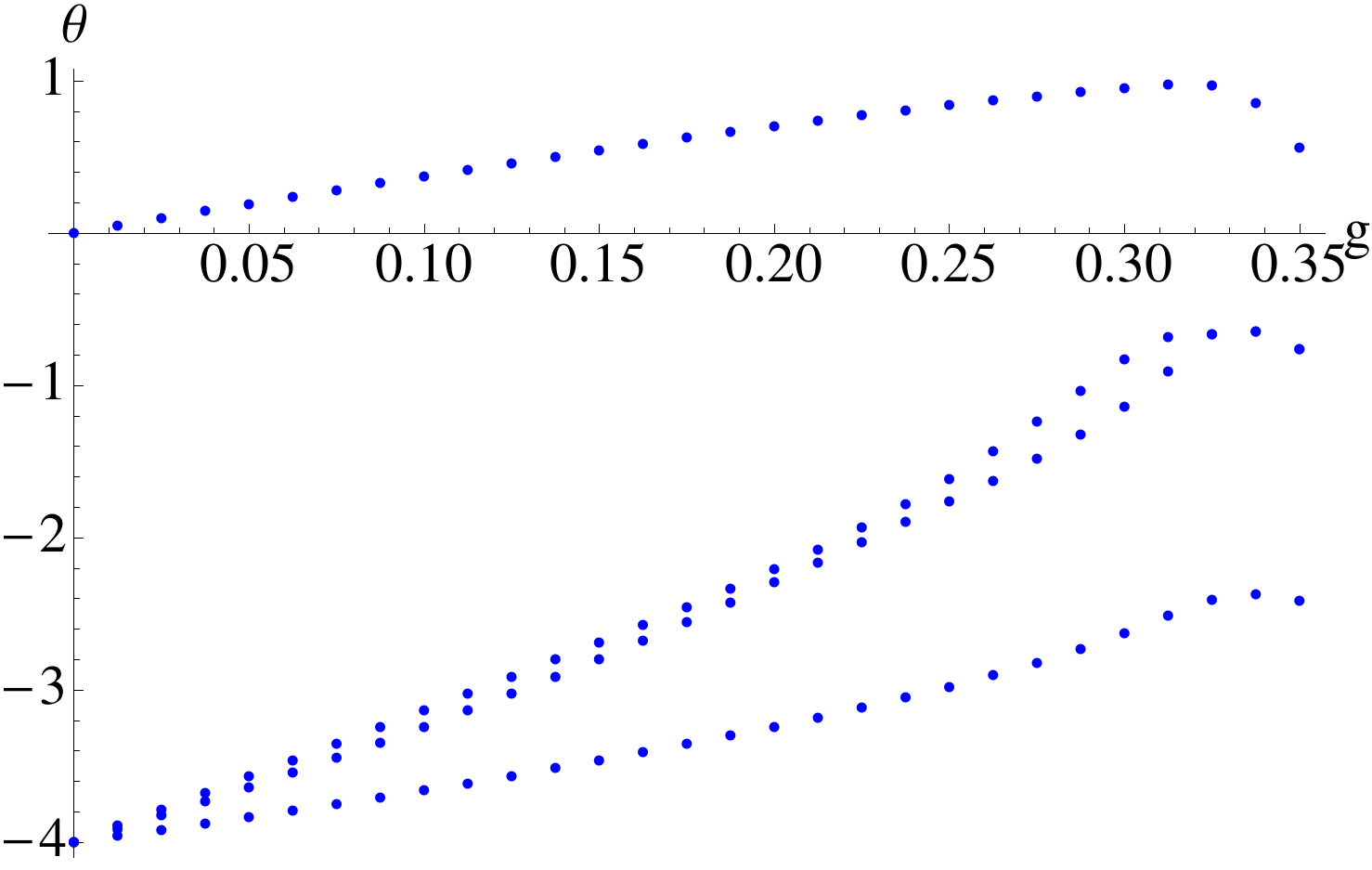}
\includegraphics[width=\linewidth]{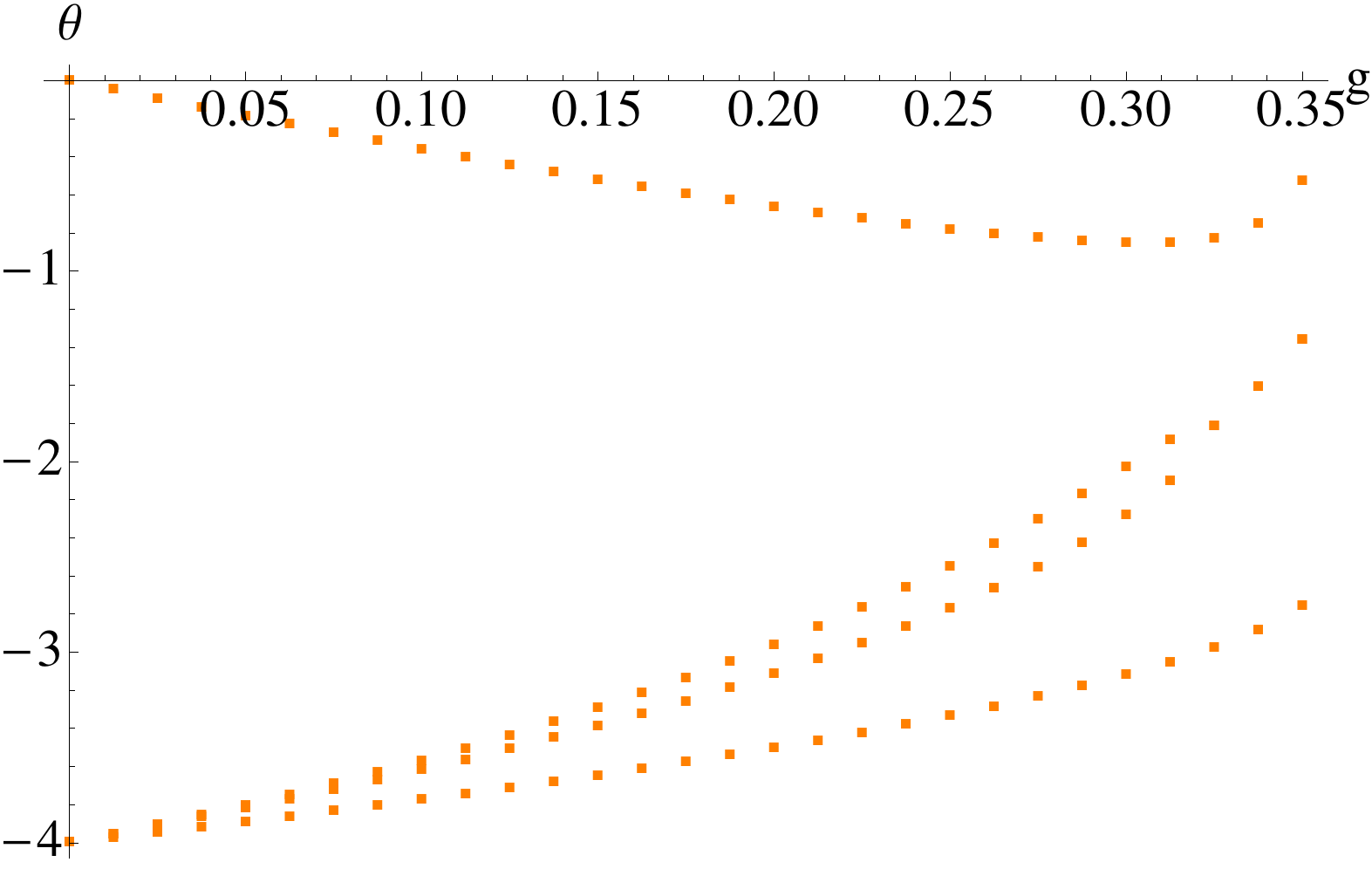}
\caption{\label{induced_FP_theta} Here we plot the four critical exponents at the two fixed points, for $\lambda=1/10$ and as a function of $g$. Clearly, at $g=0$, the critical exponents are given by the canonical dimensionality of the couplings $\bar{\rho}$ and $\lambda$. At $g>0$, they deviate significantly. Most importantly, there is one relevant critical exponent at the first fixed point (upper panel).}
\end{figure}

The situation changes at $g>0$: As is clearly visible in fig.~\ref{induced_FP}, the fixed-point values increase in their absolute value as a function of $g$. Similarly, the anomalous contributions to the critical exponents increases with $g$, see fig.~\ref{induced_FP_theta}.

At $g\neq 0$ there is no fixed point with vanishing matter couplings. We conclude that the so-called Gau\ss{}ian matter fixed point is shifted and becomes a fully non-Gau\ss{}ian fixed point, where both gravitational and matter couplings are non-vanishing. 
This confirms our claim that metric fluctuations generically induce matter self-interactions even if a free matter theory is coupled to gravity.

The mechanism behind this is simple: The most important contribution to the $\beta$ functions of the $\rho_i$ comes from diagrams, where all vertices are $\sim Z_{\phi}$. These generate a contribution to the $\beta$ functions for $\rho_i$, which is non-vanishing if all matter self-interactions are set to zero. 
Thereby the $\beta$ functions in \Eqref{betarhoschem} will have a non-vanishing contribution with $c_1 \neq 0$.
 When taking $\rho_i\rightarrow 0$, this term remains. Thus $\rho_i=0$ and $\lambda_{\phi}=0$ is \emph{not} a zero of the $\beta$ functions and thus does not correspond to a fixed point.

Most importantly, we note that the critical exponents in the matter sector deviate very significantly from their canonical values, which they would have at a Gau\ss{}ian fixed point, see fig.~\ref{induced_FP_theta}. Thus metric fluctuations not only induce a shift in the $\beta$ functions of matter couplings, such that these are non-vanishing at the fixed point, most importantly they considerably alter the scaling behaviour. This change in the scaling is due to diagrams such as a tadpole diagram and further mixed metric-matter diagrams which yield contributions $\sim g \, \rho_i$ and $\sim g\, \lambda_{\phi}$ and thus change the critical exponents in comparison to the free theory. A further contribution arises from the purely scalar two-vertex diagram, which yields contribution $\sim \rho_i \rho_j$ and $\sim \rho_i \lambda_{\phi}$ to $\beta_{\rho_i}$ and $\beta_{\lambda_{\phi}}$. Once metric fluctuations induce a non-zero value for the matter couplings, these in turn contribute to change the critical exponents from the power-counting values.
We point out that such diagrams are expected to contribute to the $\beta$ functions of all matter couplings, even those which are not induced by metric fluctuations, i.e. where a contribution of the type $\sim c_1$ (see \Eqref{betarhoschem}) is absent.  Furthermore the $\rho_i$ also couple into the flow of other matter couplings.
We might thus speculate, that a similar shift in the critical exponents as observed here, could also occur for other operators. It is thus possible that power-counting marginal operators are shifted into relevance at the shifted GFP. An example of this is clearly given by the first fixed point, where one critical exponent is actually positive, see fig.~\ref{induced_FP_theta}. Furthermore we make the important observation, that at both fixed points the critical exponents are actually shifted towards larger values. We might thus conjecture that operators of canonical dimensionality -2, which we have not included in our study, such as two-momentum four-scalar operators, will also be shifted into relevance at these fixed points.

\begin{figure}[!here]
\includegraphics[width=0.66\linewidth]{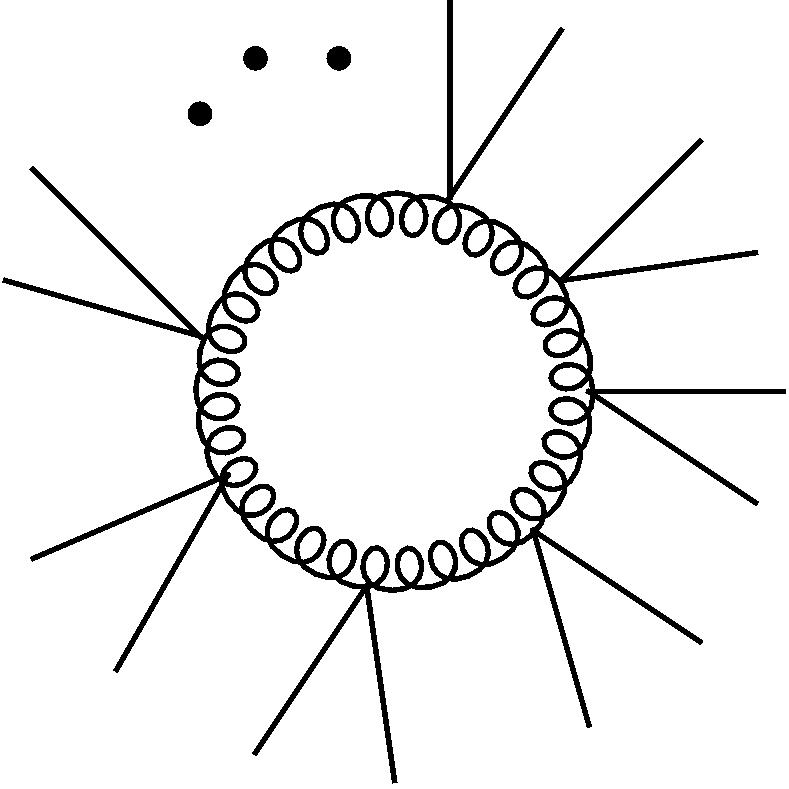}
\caption{\label{furthercouplings} Metric loops will generically induce non-vanishing scalar self-interactions with an arbitrary even number of scalar fields.}
\end{figure}

Further, our results suggest, that quantum effects will also be non-negligible for couplings beyond our truncation.
Diagrams as shown in fig.~\ref{furthercouplings} will generate further interaction terms.
In fact our calculation is a first step in a more general direction. It is directly clear from the expression for the vertices, that metric fluctuations will generate all terms of the form
\begin{equation}
\left(g^{\mu \nu}\partial_{\mu}\phi \partial_{\nu}\phi\right)^n =: \mathbf{\Phi},
\end{equation}
with $n$ integer, and similar powers of the other tensor structures. Thus the effective action will be of the form
\begin{equation}
\Gamma_{k\, \phi}= V_k(\mathbf{\Phi})\label{indaction}+...
\end{equation}
with some function $V_k$, which we have assumed to be expandable in a Taylor series and have studied the first two coefficients in the Taylor expansion here. Further terms with higher powers of momenta can also exist.

A full study of scalar fields coupled to gravity should include the full momentum-dependent potential $V(\mathbf{\Phi})$.
Let us comment on two related issues, namely the question of unitarity, i.e. absence of unphysical degrees of freedom, and the question if the effective equation of motion for $\phi$, which can be derived from \Eqref{indaction}, will allow to specify initial data with only a finite number of free parameters. As we have explained, one should generically expect that infinitely many terms of the form $\mathbf{\Phi}^n$ will acquire a non-zero coupling through metric fluctuations. Accordingly the effective equations of motion will generically contain infinitely many derivatives, thus naively requiring an infinite number of initial conditions. This naively corresponds to a loss of predictivity at the level of the effective equations of motion\footnote{There are two potential sources of loss of predictivity in a scenario with a non-Gau\ss{}ian fixed point: The first is, that infinitely many non-zero couplings only lead to a predictive theory if only a finite subset of them is relevant. This seems to be the case in asymptotically safe quantum gravity, and implies, that it is possible to write down effective equations of motion which contain an infinite number of terms, but just a finite number of free parameters. Having arrived at this step, the next question is if these effective equations of motion permit a solution with only a finite number of initial conditions, else predictivity is lost at this step.}.
As explained in detail in \cite{Barnaby:2007ve}, differential equations with an infinite number of derivatives do not generically allow to freely specify an infinite number of initial conditions. In fact, each pole in the propagator comes with 2 initial conditions, at least in the case of a free field theory. Thus the question of predictivity is closely linked to the existence of unphysical degrees of freedom: Whenever the propagator acquires a non-physical pole, further initial conditions are necessary to specify the solution to the equations of motion in the free field theory case.\\
If asymptotically safe quantum gravity is a physically permissable theory in the sense of being ghost free, then $\Gamma_{k\rightarrow p}(p^2)$ must not have a ghost pole, see \cite{Benedetti:2009rx} for indications that RG trajectories of this type exist. The precise relation between ghost poles and initial conditions must be examined in the fully interacting theory, but assuming that the analysis of the free field case carries over, the avoidance of further propagator poles at the same time implies that the solution to the equations of motion will not require infinitely many initial conditions \cite{Barnaby:2007ve}, thus the theory is predictive. 

Let us emphasise that this question of predictivity cannot be addressed from our calculation, as we truncate the effective action at order $\phi^4$.

It is interesting to observe that the fact that \emph{infinitely many} couplings are non-zero at the fixed point, which at a first glance seems to constitute a \emph{problem}, really might be the mechanism to precisely avoid this problem and render the theory ghost-free.\newline\\
Let us now consider the anomalous dimension $\eta_{\phi}$. It is of interest for several reasons, the first being, that it enters the scaling relation of any scalar coupling in the vicinity of the fixed point. To see this, consider any coupling $g_i \int d^4 x \sqrt{g} \mathcal{O}_n$, where $\mathcal{O}_n$ is an operator containing $n$ powers of the scalar field, an arbitrary power of derivatives and any further operators that depend on the metric, such as, e.g. the curvature scalar $R$. Redefining the scalar field such that the kinetic term has standard normalisation, $\phi \rightarrow \frac{\phi}{Z_{\phi}^{\frac{1}{2}}}$, yields a renormalised coupling $\hat{g}_n= \frac{g_n}{Z_{\phi}^{\frac{n}{2}}}$. Accordingly the $\beta$ function for $\hat{g}_n$ contains a term of the form $\beta_{\hat{g}} = n \eta_{\phi} \hat{g}_n+...$. To calculate the critical exponent associated with this coupling (for simplicity we assume that the stability matrix is already diagonal, but our argument carries through to the case where different operators mix) we take the derivative $-\frac{\partial \beta_{\hat{g}_n}}{\partial \hat{g}_n}=- n \eta_{\phi}+...$. 
Thus we conclude, that the question, if a coupling is relevant, depends on the anomalous dimension of the corresponding field, see also \cite{Eichhorn:2010tb} for the application of this argument to the Faddeev-Popov ghost sector. To obtain the complete set of relevant directions we need to consider the anomalous dimension.

\begin{figure}
\includegraphics[width=\linewidth]{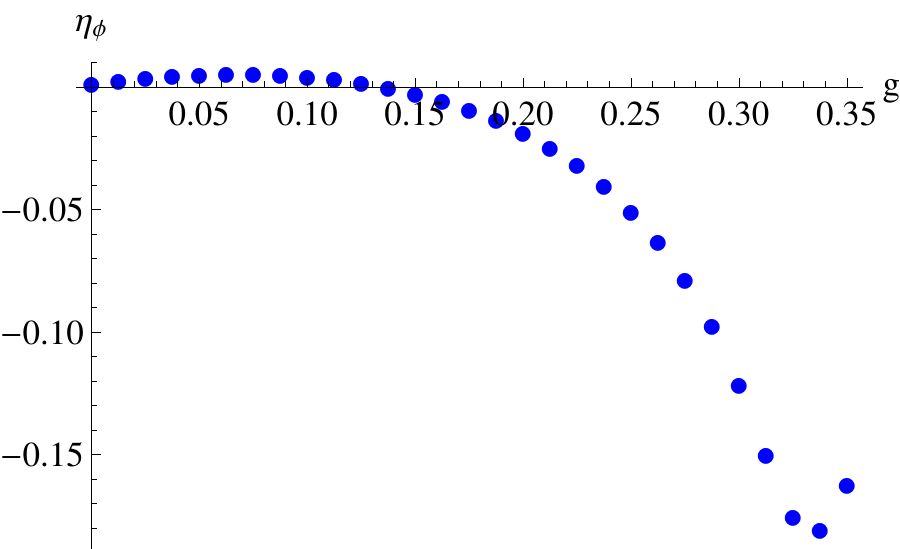}
\includegraphics[width=\linewidth]{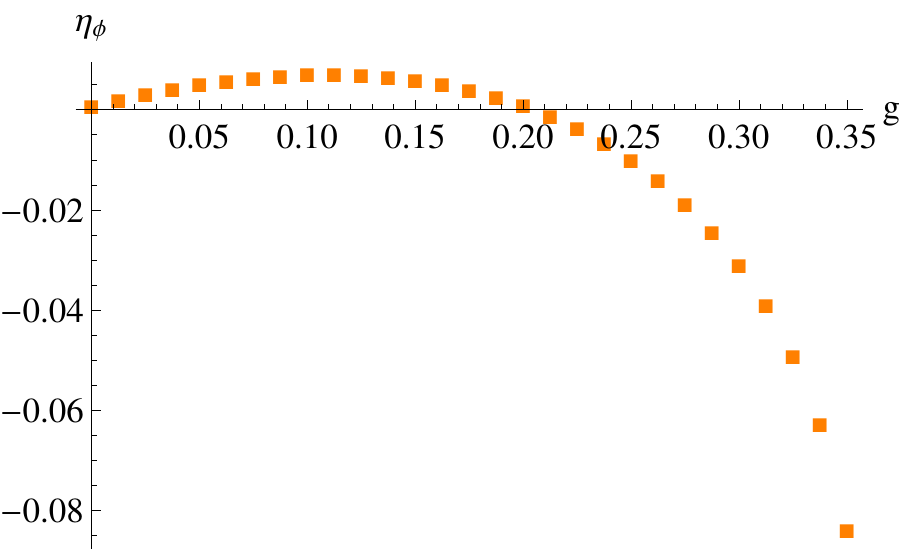}
\caption{\label{etaphiplots} Here we plot the value of $\eta_{\phi}$ as a function of $g$ for $\lambda= 1/10$ at the fixed points as in fig.~ \ref{induced_FP}. Towards larger values of $g$, $\eta_{\phi}$ becomes increasingly negative at both fixed points, implying that further matter couplings could be shifted into relevance.}
\end{figure}

In the case of the scalar field coupled to gravity, we observe that increasing values of $g$ induce a change of sign in $\eta_{\phi}$ at both fixed points, such that $\eta_{\phi}$ becomes negative, see fig.~\ref{etaphiplots}.
Thus further matter couplings besides those included in our truncation, may become relevant. 

The anomalous dimension is also of interest for a second reason: Asymptotic safety is assumed to yield a fractal spacetime in the sense that one manifold is endowed with a family of metrics, labelled by the RG scale $k$ \cite{Lauscher:2005qz,Reuter:2011ah,Reuter:2012id}. In particular, indications are found that in the fixed-point regime, an effective dimensional reduction to $d=2$ occurs. 
There are several possible interpretations for this result: The first implies that the effective dimensional reduction is visible to all fields, i.e. metric as well as matter fluctuations behave as in 2 dimensions. The second interpretation is, that this effective dimensional reduction is felt by the gravitational degrees of freedom only, but matter fields cannot be described as propagating in 2 dimensions at high momentum scales. 
The first interpretation would require $\eta \rightarrow -2$ for all fields, since then the propagator scales logarithmically with distance, i.e. as in 2 dimensions. Naturally, we do not expect to obtain the exact value $-2$ in any truncation. On the other hand we do expect, that the sign of $\eta$ is not dependent on the truncation, since the sign crucially decides on the number of relevant operators, which is generally expected to be a truncation-independent quantity in the sense that the relevance of any operator should not depend on the truncation in which its $\beta$ function is evaluated. 
Thus a positive anomalous dimension would imply, that the dimensional reduction is a mechanism which applies to the propagation of gravitational fluctuations, but not necessarily to matter fluctuations.

Here, we observe that for increasing $g$, the values of $\eta_{\phi}$ become negative. Extensions of the truncation can induce changes $\mathcal{O}(1)$ in these quantities, so whether a values of $\eta_{\phi} \sim -2$ will be approached in a more complete truncation remains to be investigated. We conclude that the observed value of $\eta_{\phi}$ at the shifted Gau\ss{}ian fixed point in this truncation is not incompatible with some form of dimensional reduction.

A crucial consequence of the anomalous dimension is the momentum-dependence of the propagator, $\left(\Gamma_k^{(2)}\right)^{-1}= \left( Z_k p^2\right)^{-1}$. In a single-scale setting, where the momentum of the scalar field sets the physical scale, the identification $k^2=p^2$ is reasonable, its justification being that in such a setting a tree level evaluation of the effective action suffices to evaluate the leading order contribution to physical quantities such as scattering cross sections. Thus we have that $\left(\Gamma_k^{(2)}\right)^{-1}=\left( p^2\right)^{1-\eta/2}$. 
This type of RG improvement can be used to deduce experimental consequences of asymptotically safe quantum gravity (or, within the effective field-theory framework, of other UV completions for gravity). Here we conclude that the observed values of $\eta_{\phi}<0$ for larger $g$ imply a suppression of scalar fluctuations in comparison to the perturbative expectation. Thus, the contribution of high-momentum scalar fluctuations in loop diagrams is suppressed, which can have observable consequences.
Note that $\eta_{\phi}>0$ makes the scalar propagator even more divergent than in perturbation theory. However this does not imply that physical quantities will be divergent at this fixed point. It is to be expected that anomalous scaling of vertices yields finite values for scattering amplitudes etc.\\

\subsection{Fixed-point analysis for the full truncation}\label{fulltruncation}
As discussed, we observe two interating fixed points in the matter sector at small values of $g$ and $\lambda$, which are connected continuously to the Gau\ss{}ian fixed point. These vanish if we go beyond certain critical values of $g$ and $\lambda$, e.g. along the line $\lambda= 1/10$, they vanish for $g>0.35$. Let us note that within an extended truncation, this can very well change: On the one hand, further contributions to the matter beta functions exist, e.g. from further matter couplings, such as $\phi^6$ couplings. These can in principle balance the effect of metric fluctuations and allow the shifted Gau\ss{}ian fixed points to exist beyond the values found in this truncation. Furthermore, the fixed-point values in the gravitational sector depend on the inclusion of further operators, and further matter fields. Finally, we work within a single-metric approximation, and do not take the bimetric structure of the RG flow in the background-field formalism into account in our calculation.
Thus our results do not imply that within extended truncations, a shifted Gau\ss{}ian fixed point cannot exist in the full truncation.

Here we make the following observation: The full system of $\beta$ functions for the matter and gravity sector (see app.~\ref{matterbetaunspecifiedr} for the matter $\beta$ functions for general shape function) does not admit a Gau\ss{}ian matter fixed point (where gravitational couplings are non-zero, but matter couplings are zero), and it also does not admit the existence of an interacting fixed point which is connected continuously to the Gau\ss{}ian one for $g \rightarrow 0$. 

We find a non-Gau\ss{}ian fixed points at considerably larger values of the matter couplings, see tab.~\ref{fptable}.
This fixed point seems to be the continuation of the third fixed point in the pure matter system, see tab.~\ref{fptablewograv}. Note that the effect of metric fluctuations is to considerably change the values of the critical exponents, albeit keeping them negative thus corresponding to irrelevant couplings.

\begin{widetext}

\begin{table}[!here]
\begin{tabular}{c|c|c|c|c|c|c|c|c|c|c}
$g_{\ast}$& $\lambda_{\ast}$& $\rho_{a\, \ast}$& $\rho_{b\, \ast}$& $\rho_{c\, \ast}$& $\lambda_{\phi\, \ast}$& $\eta_{\phi\, \ast}$& $\theta_{1,2}$& $\theta_{3,4}$& $\theta_{5}$& $\theta_{6}$\\
\hline
0.264 & 0.391 & 47.094&24.771&66.120&-127.502&2.718&2.122$\pm i$ 1.716 &- 0.160$\pm i$19.242& -12.840&-6.245
\end{tabular}
\caption{\label{fptable} Here we give the fixed-point values of all, gravitational and matter couplings, at the non-Gau\ss{}ian fixed point. We also list the value of the anomalous dimension, and the critical exponents. We find two critical exponents with positive real part, as well as for critical exponents with negative real part, corresponding to to irrelevant directions.}
\end{table}

\end{widetext}

Since the momentum-independent coupling $\lambda_{\phi}$ is negative at the fixed point, the Euclidean effective potential seems unbounded from below. Here it is important to realise that higher-order terms in the scalar field will be induced by metric and scalar fluctuations. These can result in a bounded effective potential, even for $\lambda_{\phi}<0$. Thus we cannot discard this fixed point on the grounds of stability.

The mechanism, why the shifted Gau\ss{}ian fixed point in the matter couplings vanishes completely for larger values of $g$ and $\lambda$, but not the non-Gau\ss{}ian one, is simple to understand: Whereas the effect of metric fluctuations is to shift the matter $\beta$ function in such a direction as to induce the vanishing of fixed points, matter fluctuations have the opposite effect: The pure matter system admits the existence of non-trivial fixed points. Thus the only scenario in which a full fixed point can survive the onset of metric fluctuation is one where the matter sector itself is strongly interacting, since then clearly the effect of metric fluctuations is less dominant than at the Gau\ss{}ian matter fixed point.

We thus arrive at the following conclusion: In a scalar theory coupled to asymptotically safe quantum gravity, there is no Gau\ss{}ian matter fixed point. For very small values of $g$ it becomes shifted into an interacting fixed point. Finally, at larger values of $g$, where the gravitational $\beta$ functions have fixed points, the shifted Gau\ss{}ian fixed point in the matter sector does not exist any more (i.e. the zeros of the $\beta$ function lie in the complex plane away from the real axis). There is a fully non-Gau\ss{}ian fixed point at larger values of the matter couplings. Canonical power-counting does not hold at this fixed point, due to the large fixed-point values.

We conclude that in constructing a quantum theory of gravitational and matter degrees of freedom based on asymptotic safety the fact that the gravitational sector remains interacting in the UV implies that there will be a strongly-coupled matter sector. Within our truncation we find no evidence for a picture where the far UV is dominated by metric fluctuations only, and the matter sector becomes trivial. In contrast, the UV behaviour of the theory is determined by an interacting matter and gravity theory, where metric fluctuations induce non-vanishing matter couplings, and accordingly matter fluctuations become important and drive the running couplings in both sectors.

Interestingly, none of the matter couplings is shifted into relevance, although quantum fluctuations induce large departures from canonical scaling. This is also reflected in the large anomalous dimension $\eta_{\phi}$. 
Note however, that two of the critical exponents are rather close to zero. It is thus conceivable, that similar shifts in couplings of canonical dimensionality $-2$ result in these couplings becoming relevant. Therefore, extended truncations of scalar matter coupled to gravity might exhibit more relevant couplings than the sum of relevant couplings in the gravitational and the matter sector considered separately.

We observe that since this fixed point is dominated by matter fluctuations, the anomalous dimension for the scalar is positive. 

Let us discuss the dependence of our results on the choice of regulator: Here, we employ a spectrally adjusted cutoff, that implies the existence of terms $\sim \eta_{\phi}, \eta_N, \partial_t \lambda$ on the right-hand side of the flow-equation. These terms result in a clearly non-perturbative structure of the flow equation, since $\eta_{\phi}$ will also depend on the inverse of the matter couplings, see app.~\ref{anomdimapp}. In Yang-Mills theory, this structure is crucial to uncover the existence of an infrared attractive non-Gau\ss{}ian fixed point, see \cite{Gies:2002af}.\\
A regulator of this type has also been employed in \cite{Eichhorn:2010tb} to evaluate the ghost anomalous dimension in asymptotic safety, as compared to a non-spectrally adjusted one for the evaluation of the same quantity in \cite{Groh:2010ta}. In the same choice of gauge, the difference in regularisation scheme yielded $\eta_c = -1.3$ vs. $\eta_c = -1.8$. This exemplifies that numerical differences follow from the use of one versus the other regularisation scheme, but important conclusions, in particular about the relevance of couplings, are not affected. Most importantly, as has been pointed out in \cite{Narain:2009qa}, there exists a choice of non-spectrally adjusted cutoff which reproduces the fixed-point values found with a spectrally adjusted cutoff. We thus conclude that our main results on the existence of fixed points and most importantly the critical exponents should not be affected by a change in the cutoff procedure.

Since new matter-couplings are induced by metric fluctuations, it is interesting to investigate their backreaction onto the gravitational sector. This is another test of the consistency of the asymptotic-safety scenario:
To decide whether a fixed point is an artifact of a truncation, or exists in full theory space, it is useful to investigate its stability under extensions of the truncation. Furthermore, the numerical values of the critical exponents are a good measure of the stability of the fixed point under such extensions. Numerous results exist showing the stability of the fixed point under various extensions of the truncation \cite{Lauscher:2001ya,Reuter:2001ag,Lauscher:2002sq,Percacci:2002ie,Percacci:2003jz,Litim:2003vp,Fischer:2006fz,Codello:2008vh,Benedetti:2009rx,
Zanusso:2009bs,Vacca:2010mj,Groh:2010ta,Eichhorn:2010tb,Manrique:2010am,Manrique:2011jc,Donkin:2012ud}. Here, we add further evidence, by including the couplings $Z_{\phi}$ and $\rho_i$ into the flow of $g$ and $\lambda$.
At a first glance, one would not expect a backcoupling of $\rho_i$ into the gravitational $\beta$ functions $\beta_g$ and $\beta_{\lambda}$, since the terms $\sim \rho_i$ in the effective action do not contribute to the scalar two-point function at vanishing scalar field, see also sect.~\ref{calculation}.
However the $\rho_i$ contribute to $\eta_{\phi}$, which does of course enter the gravitational $\beta$ functions. Since $Z_{\phi}(k)$ corresponds to an inessential coupling, it can be eliminated from the other $\beta$ functions, by simply inserting $\eta_{\phi}(g, \lambda, \rho_a,\rho_b,\rho_c)$. In this fashion the $\rho_i$ enter $\beta_{g}$ and $\beta_{\lambda}$, and can thus alter the RG flow of $g$ and $\lambda$.

\begin{figure}[!here]
\includegraphics[width=\linewidth]{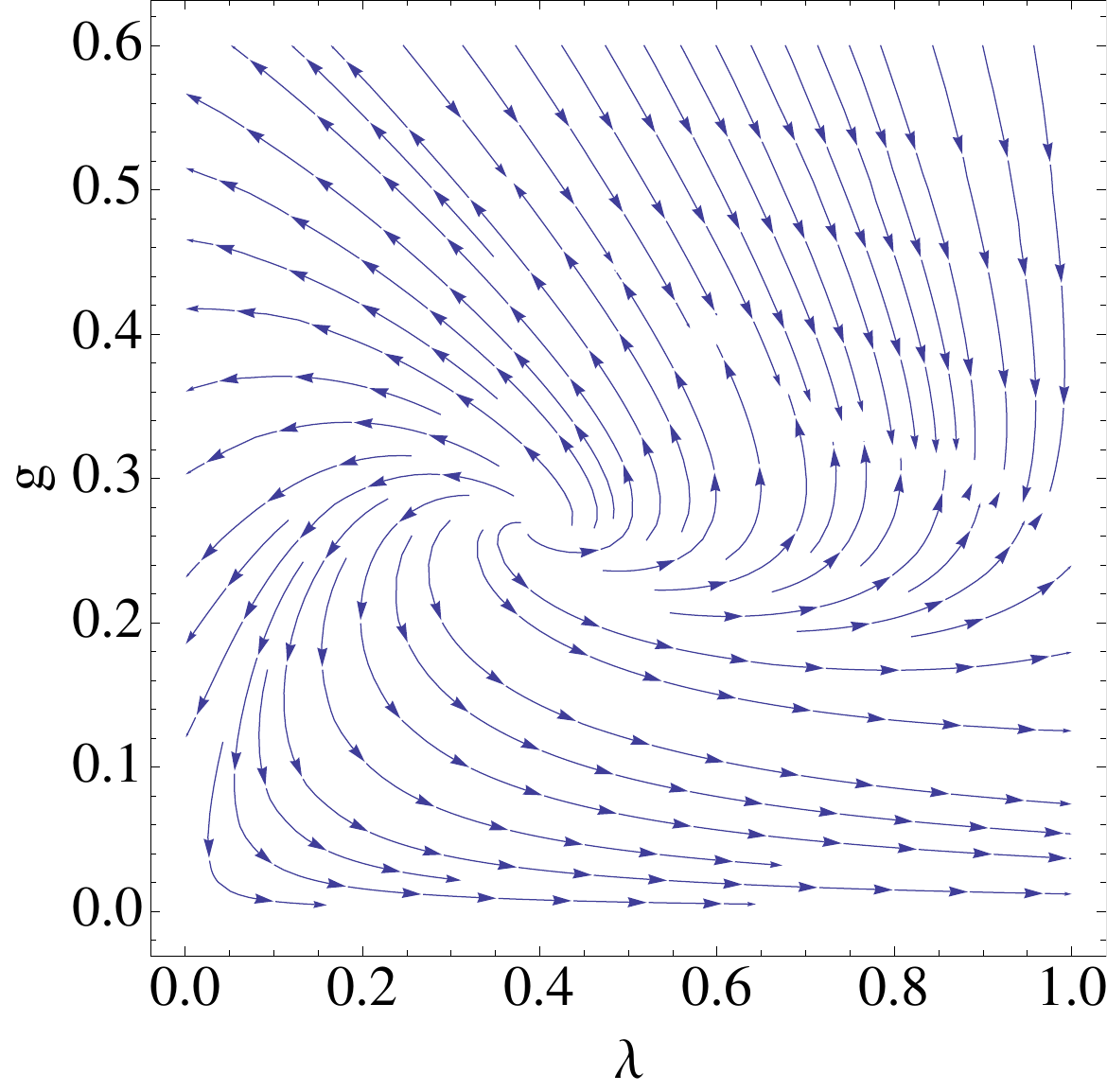}
\caption{\label{EHplot} Here we plot the RG flow towards the infrared, projected onto the Einstein-Hilbert plane for $\rho_a=47.094$, $\rho_b=24.771$, $\rho_c=66.120$ and $\lambda_{\phi}= -127.502$, cf. tab.~\ref{fptable}. Clearly the flow resembles the flow in the pure Einstein-Hilbert truncation at vanishing matter couplings to a very high degree and admits trajectories passing very close to the Gau\ss{}ian fixed point and yielding a constant value of the dimensionful Newton coupling and cosmological constant, in  accordance with observations.}
\end{figure}

As a result, we point out that the RG flow projected onto the Einstein-Hilbert subsector $\{\beta_g, \beta_{\lambda}\}$ is very stable under this extension of the truncation. As can be seen in fig.~\ref{EHplot}, changing the value of the matter couplings considerably only leads to a very slight change in the fixed-point values and the structure of the RG flow in the Einstein-Hilbert sector.

 In particular, our extended truncation again permits to choose initial conditions for the RG flow within the UV-critical surface of the NGFP which give a trajectory with a long classical regime, see fig.~\ref{EHplot}, and, e.g. \cite{Reuter:2004nx}.

Taken together, we take this as further evidence for the NGFP in the gravity theory \emph{not} being an artifact of the truncation, but existing in full theory space.

\section{The possibility of observable effects in the CMB}

Inflation with a scalar field within asymptotically safe quantum gravity has been studied in \cite{Contillo:2011ag}.
The main conclusion is that inflation and a scale-invariant spectrum of scalar perturbations are possible in such a setting.

As is known, derivative interactions do not alter the slow-roll conditions for inflation, since within the standard scenario, the classical background value of the scalar field is constant over space, and $\frac{1}{2}\dot{\phi}^2<< V(\phi)$. Clearly adding any type of derivative-interaction to the potential respects these conditions. Momentum-dependent interaction terms have been studied in the context of inflation, see, e.g. \cite{Burrage:2010cu}.
Thus our result, that metric fluctuations generate momentum-dependent scalar interactions, will not affect the conclusion that asymptotically safe quantum gravity can admit inflation with a scalar field.

Interestingly, one of the operators that we investigate here, namely $\Phi^2$ has been shown \cite{Creminelli:2003iq} to potentially give larger non-Gau\ss{}ianities in the spectrum of the scalar field fluctuations than in the standard slow-roll inflation without this additional term \cite{Maldacena:2002vr}. 

Following \cite{Creminelli:2003iq}, the result for the non-Gau\ss{}ianity parameter $f_{NL}$\footnote{This parameter is defined as follows: The power spectrum $\mathcal{P}(p_1, p_2)$ and the bispectrum $\mathcal{B}(p_1, p_2,p_3)$ of the scalar perturbations are defined via the Fouriertransform of the two- and the three-point function, modulo a $\delta$ function for momentum-conservation. Then the simplest measure for non-Gau\ss{}ianity is $f_{NL}= \frac{5}{18}\frac{\mathcal{B}(p,p,p)}{\mathcal{P}(p,p)^2}$.} is $f_{NL} \sim \bar{\rho}_a$. The value for $f_{NL}$ thus depends on the identification of $k$: Evaluating the three-point function of the scalar field fluctuations during inflation requires the identification of the RG scale $k$ with a physical scale of the system. Here, a possible choice is $k \sim H$ where $H$ is the Hubble scale during inflation, see, e.g. \cite{Bonanno:2009nj} and references therein. The precise scale-identification is crucial for a quantitatively meaningful prediction of the non-Gaussianity. 

Since asymptotically safe quantum gravity necessitates the existence of such a term, this effect could in principle be an observable consequence of asymptotic safety with experiments.\\

For a prediction of the numerical value of $f_{NL}$, the flow of $\bar{\rho}_a$ on a trajectory potentially describing our universe is necessary. In particular this will crucially depend on terms beyond our truncation, such as $\mathcal{O}(\phi^6)$ terms, which give a direct contribution to $\beta_{\rho_a}$. 

In a more general setting, fluctuations of the metric generate this term within the effective-field theory setting. Since it leaves an observable imprint in the CMB, measurements of the non-Gau\ss{}ianity allow us in principle to access the RG flow of Newton's coupling and the cosmological constant in other quantum gravity proposals. In more detail, the idea is the following one: Different quantum gravity proposals differ in the choice of their fundamental variables, the realisation of fundamental symmetries etc. Thus the "pure quantum gravity regime", in which even the notion of a continuous spacetime is often postulated to break down, looks widely different in distinct quantum gravity proposals.
They agree on the necessity to recover classical Einstein gravity at small momentum (large length) scales. Presumably there is a regime between a "pure quantum gravity regime" and the classical regime, in which quantum gravity fluctuations start to be important, but they can be calculated within a framework, in which the symmetries of Einstein gravity hold, and the metric is used as an effective degree of freedom. This regime is a semiclassical regime, where our calculation holds, and it presumably dominates energy scales at which inflation takes place. Within this regime, the RG framework can be applied. The crucial point to realise is that different proposals for quantum gravity will translate into different initial (i.e. high-energy) values for the couplings, at which the RG flow will start. In order for the quantum gravity proposal to be realised in our universe, the RG flow must reach a region where $G_N = \rm const$ on small momentum scales. On scales slightly below the initial UV scale, the values of different couplings can be very different for different quantum gravity proposals. What they all have in common is the generation of matter-interaction terms such as the one that we have investigated here. Thus many proposals for quantum gravity, different as they may be in the very far UV, will all generate a non-zero value for $\rho_a$ at energy scales presumably below the Planck scales. Bridging the gap between the fundamental description and the effective-field theory description then allows to predict as specific RG trajectory $\rho_a(k)$ for a specific UV completion. As $\rho_a$ is in principle accessible to observations in the non-Gau\ss{}ianity of the CMB, we can in principle use this observation to infer the RG trajectory of the Newton coupling and the cosmological constant. Thus different UV completions for gravity can be distinguished at the level of observation.

\section{Conclusions and outlook}

We have studied quantum gravity coupled to matter, and shown that in the case of a scalar field, quantum gravity fluctuations, which we parameterise as metric fluctuations, generate scalar self-interactions when coupled to gravity.
We point out that this is a generic feature of quantum gravity fluctuations: Whenever a free matter field theory is coupled to gravity, the $\sqrt{g}$ generates matter-graviton vertices, and  graviton loops generate matter self-interactions, in particular in the non-perturbative regime.
Here we have shown that in the case of a scalar field, the induced interactions are momentum dependent, and thus have the form $\left(\partial_{\mu}\phi \partial^{\mu}\phi\right)^{2n}$, $\phi^n (\nabla^2 \phi)^n$ and $\left(\partial_{\mu}\phi \partial^{\mu}\phi\right)^n (\phi \nabla^2 \phi)^n$ with $n \geq 1$. Most importantly, these matter self-interactions also couple into the flow of the scalar effective potential, and remove the Gau\ss{}ian fixed point in the scalar momentum-independent coupling $\lambda_{\phi}$.
Thus we conclude that quantum gravity fluctuations alter the properties and the dynamics of matter systems.\\
Let us stress that this result holds within asymptotically safe quantum gravity, but also in an effective-field-theory setting, where the UV completion for gravity is not specified. We thus conclude, that quantum gravity effects generically alter the properties and dynamics of matter theories in comparison to the gravity-free case.

As shown in a previous work \cite{Eichhorn:2011pc,Eichhorn:2011ec}, a similar mechanism is at work when fermions are coupled minimally to gravity: Starting with only a minimally coupled kinetic term, four-fermion interactions are induced, even if these four-fermion couplings are set to zero initially. Due to the dimensionality of $\frac{3}{2}$ of a fermion field, the induced fermionic self-interactions are not momentum-dependent in this case.

Let us comment on the interesting case of gauge fields coupled to gravity, studied in \cite{Robinson:2005fj,Pietrykowski:2006xy,Toms:2007sk,Ebert:2007gf,Tang:2008ah,Daum:2009dn,Daum:2010bc,Folkerts:2011jz}, which is the subject of future work: In a similar way to what we observed here, higher-order gauge-field self-interactions will presumably be generated. Since there is no symmetry to forbid the generation of these terms, only a highly non-trivial cancellation mechanism between diagrams could avoid the generation of some of the infinitely many $\left(F_{\mu \nu}F^{\mu \nu}\right)^n$ terms. In principle, quantum gravity effects could even lead to an unstable matter action, thus excluding this type of UV completion for gravity.

Let us also note that a similar mechanism is expected to be at work in a first-order formulation of gravity, as examined in \cite{Daum:2010qt,Harst:2012ni} in the context of asymptotic safety.

Noting that the metric-induced matter interactions can alter the dynamics of matter fields at high momenta, and also potentially increase the number of relevant couplings, and thus the free parameters of the theory, one might wonder if there is a way to avoid this mechanism. At a first glance, unimodular quantum gravity \cite{Weinberg:1988cp,Smolin:2009ti}, where $\sqrt{g}$ is held fixed, and which, at the classical level is equivalent to General Relativity, might provide a solution: One might think that holding $\sqrt{g}$ fixed avoids the generation of matter-graviton-interaction vertices from a free theory. However this is only partially right, since kinetic terms usually contain further factors of the metric. Even if the covariant derivative reduces to a partial derivative, as in the case of a scalar field, further factors of the inverse metric exist (as in $g^{\mu \nu} \partial_{\mu}\phi \partial_{\nu}\phi$). These can be expanded in an infinite series in the fluctuation metric, thus again generating metric-matter interactions, which induce matter self-interactions through loop diagrams.

We conclude that the mechanism which generates matter self-interactions when a free matter theory is coupled to gravity, seems to be rather generic, and can presumably not be avoided for any type of matter, be it scalar, fermionic or spin-1.

Within our truncation, we find the following interesting results: Metric fluctuations shift the so-called Gau\ss{}ian matter fixed point to make it an interacting one, where momentum-dependent interaction terms have non-zero couplings and also induce a non-vanishing value for the momentum-independent coupling $\lambda_{\phi}$. First treating the (dimensionless) Newton coupling $g$ and the cosmological constant $\lambda$ as external parameters, we find two fixed points in the matter $\beta$ functions which are connected continuously to the Gau\ss{}ian fixed point for $g \rightarrow 0$ and which exist for small values of $\lambda$ and $g$.
Most importantly, the critical exponent of the matter couplings at the fixed points are changed considerably in comparison to the Gau\ss{}ian case, and in fact $\lambda_{\phi}$ can turn into a relevant coupling.

Studying the full system of $\beta$ functions for $g, \lambda, \rho_{a}, \rho_{b}, \rho_c, \lambda_{\phi}, Z_{\phi}$ we note that the shfited Gau\ss{}ian fixed points cease to exist: They exist for small values of $g$, but since metric fluctuations have the effect shifting the matter $\beta$ functions such that fixed points tend to be removed, values around $\{g \sim 0.25, \lambda \sim 0.35\}$ are too big in order for the shifted Gau\ss{}ian fixed point to exist. In the full truncation, only a non-Gau\ss{}ian fixed point with large values of the matter couplings exist. In the limit $g \rightarrow 0$, where metric fluctuations are turned off, it is not continuously connected to the Gau\ss{}ian fixed point.
At this non-Gau\ss{}ian fixed points, the matter couplings remain irrelevant, but the critical exponents deviate from the power-counting dimensionality of the couplings. The gravitational couplings correspond to relevant directions. Thus our 6-coupling theory space admits a UV completion which is fully interacting and has two free parameters.

We therefore conclude that within asymptotically safe quantum gravity, it is crucial to take into account the metric-induced matter self-interactions, since their existence implies that in the far UV, both gravitational as well as matter couplings are non-zero and therefore both sector contribute to the RG flow. 
Most importantly, the critical exponents deviate significantly from the canonical scaling dimensions. We thus conclude, that within a more extended truncation, matter couplings with negative canonical dimensionality could even be shifted into relevance.

Note furthermore that matter self-interactions generically couple back into the flow of the gravitational couplings. In our case, this happens since the anomalous dimension $\eta_{\phi}$, which couples into $\beta_g$ and $\beta_{\lambda}$, depends on the $\rho_i$. Potentially, gravity-induced matter self-interactions can then crucially alter the RG flow in the gravitational sector. In the truncation investigated here, the change of the gravitational RG flow in comparison to the matter-free case is only mild, but a complete study of the gravitational RG flow must take into account the metric-induced matter self-interactions.

Future directions of this work should obviously include extension of the truncation in the matter sector, but also resolve the single-metric approximation and account for the bimetric nature of the RG flow in the backgrounf-field formalism in gravity. Finally, further matter fields, that in more realistic toy models of the standard model obviously couple to both gravity as well as the scalar sector, can also have a crucial effect on the fixed-point structure in the UV.

Note also that the mechanism that we explained here, also applies to Faddeev-Popov ghosts fields, which arise in the context of gauge-fixing. Metric fluctuations can be expected to generate further terms involving ghost fields beyond the simple exponential of the Faddeev-Popov determinant. It should be expected that the ghost sector of asymptotically safe quantum gravity differs significantly from a perturbative ghost sector. Whether this allows for a solution of the Gribov problem in quantum gravity remains to be investigated.

We point out that momentum-dependent scalar interaction terms as the one investigated here have been examined in the context of inflation: Assuming that the slow-roll conditions for a scalar field are satisfied, this type of term is also allowed. However it does have interesting consequences for possible observations, since it can induce larger non-Gau\ss{}ianities in the CMB than within a standard slow-roll scenario. We conclude that, although quantitative precision cannot be expected from our truncation, asymptotically safe quantum gravity might be expected to leave a potentially observable imprint in large non-Gau\ss{}ianities in the CMB, which might be observable.\newline\\

{\emph {Acknowledgements}}\\
I would like to thank H. Gies for helpful comments on the manuscript and N. Afshordi for helpful discussions.

Research at Perimeter Institute is supported by the Government of Canada through Industry Canada
and by the Province of Ontario through the Ministry of Research and Innovation.

\begin{appendix}
\section{Vertices}\label{appendix1}

In the following we list all vertices which can be derived from our truncation. Here, $h = \bar{g}^{\mu \nu}h_{\mu \nu}$ and $\bar{g}^{\mu \nu}h_{\mu \nu}^{TT}=0$ and $\bar{D}^{\mu}h_{\mu \nu}^{TT}=0$. Further components of a decomposition of the fluctuation metric do not couple into the matter $\beta$ functions. 
We follow the conventions
\begin{equation}
\Phi_a(x)= \int \frac{d^4p}{(2 \pi)^4}\Phi_a(p)e^{- i \, p \cdot x}
\end{equation}
and 
\begin{equation}
\Gamma^{(2)}_{\Phi_a \Phi_b }(p,q)= \frac{\delta^2 \Gamma_{k}}{\delta \Phi_a(p)\delta \Phi_b(-q)}.
\end{equation}
Herein $\Phi_a$ denotes the components of a superfield, thus $\Phi_a= \left(\phi, h_{\mu \nu}^{TT}, h \right)$.

\begin{widetext}
With this conventions in mind our truncation yields the following vertices:
\begin{eqnarray}
\Gamma_{k\, h \phi}^{(2)}(p,q)&=& \frac{1}{4}Z_{\phi}(k) \left(-p \cdot q +q^2\right) \phi(q-p)\nonumber\\
&{}&+ \int_{l_1,l_2} \phi(l_1)\phi(l_2)\phi (q-p-l_1-l_2)\Bigl(\frac{\bar{\rho}_b}{2}\left( - p \cdot l_1 l_1^2- p^2 l_1^2+p \cdot l_1 q^2\right) + 2 \bar{\lambda}_{\phi} \nonumber\\
&{}&+\frac{\bar{\rho}_c}{4}\left(-l_1 \cdot q\, p \cdot l_2 - l_1 \cdot l_2 p \cdot q - p \cdot l_1 p \cdot l_2 - l_1 \cdot l_1 p \cdot l_2 - l_1 \cdot l_2 p \cdot l_2\right)\Bigr).
\end{eqnarray}
Here we use the following notation: $\int_q = \int \frac{d^4 q}{(2 \pi)^4}$.
\begin{eqnarray}
\Gamma_{k\, \phi\, h}^{(2)}(p,q)&=&\frac{1}{4}Z_{\phi}(k) \left( -p\cdot q+p^2\right) \phi(q-p)\nonumber\\
&{}&+ \int_{l_1,l_2} \phi(l_1)\phi(l_2)\phi (q-p-l_1-l_2)\Bigl(\frac{\bar{\rho}_b}{2}\left( q \cdot l_1 l_1^2- q^2 l_1^2-q \cdot l_2 p^2\right) +2 \bar{\lambda}_{\phi}\nonumber\\
&{}&+\frac{\bar{\rho}_c}{4}\left(-l_1 \cdot p\, q \cdot l_2 - l_1 \cdot l_2 p \cdot q - q \cdot l_1 q \cdot l_2 + l_1 \cdot l_1 q \cdot l_2 + l_1 \cdot l_2 q \cdot l_2 \right)\Bigr).
\end{eqnarray}

We make the following useful observation: From the kinetic term, there is no vertex with two internal scalar gravitons, so the vertex is $\sim \rho_i$:
\begin{eqnarray}
\Gamma_{k\, h\, h}^{(2)}(p,q)&=&\int_{l_1,l_2,l_3} \phi(l_1)\phi(l_2)\phi(l_3)\phi(q-p-l_1-l_2-l_3)\Bigl(\frac{\bar{\rho}_b}{8}\left( q \cdot l_1 l_2^2-p \cdot l_1 l_2^2-\frac{1}{2} p \cdot l_1 q \cdot l_2 -\frac{1}{2} q \cdot l_1 p \cdot l_2\right) \nonumber\\
&{}&- \frac{\bar{\rho}_c}{16} \left(-l_1 \cdot l_2 q \cdot l_3 + l_1 \cdot l_2 p \cdot l_3\right) + \frac{1}{8}\bar{\lambda}_{\phi}\Bigr).
\end{eqnarray}
The vertices coupling the scalar to the transverse traceless graviton component read as follows:
\begin{eqnarray}
\Gamma_{k\, h_{\mu \nu}^{TT}\, \phi}(p,q)&=& -\frac{Z_{\phi}(k)}{2}\phi(q-p)\left(2 q_{\mu}q_{\nu}-q_{\mu}p_{\nu}-q_{\nu}p_{\mu}+p_{\mu}p_{\nu} \right)\nonumber\\
&+&\int_{l_1,l_2} \phi(l_1)\phi(l_2)\phi(q-p-l_1-l_2)\Bigl(-2 \bar{\rho}_b(k) \left(2l_{2\mu}l_{2\nu} l_1^2 +q_{\mu}q_{\nu}l_2^2+ l_{2\mu}l_{2\nu}q^2 \right)\nonumber\\
&{}&-2 \bar{\rho}_a(k) \left(\left(-q_{\mu}l_{1\, \nu}-q_{\nu}l_{1\, \mu}\right) (-l_2 \cdot p +l_2 \cdot q-l_1\cdot l_2 -l_2^2)+\left(l_{1\, \mu}l_{2\, \nu}+l_{1\, \nu}l_{2\, \mu} \right) (q\cdot p+q\cdot l_1+q \cdot l_2-q^2)\right) \nonumber\\
&{}& - \frac{\bar{\rho}_c}{2} \Bigl(l_{2\mu}l_{2\nu}\left(-2 l_1 \cdot q-2l_1 \cdot p-2l_1^2-2 l_1 \cdot l_2 \right)+q^2 \left(l_{1\mu}l_{2\nu}+l_{1\nu}l_{2\mu} \right)+2 q_{\mu}q_{\nu}l_1 \cdot l_2 \nonumber\\
&{}&\quad \quad + l_2^2 \left(-l_{1\mu}q_{\nu}-l_{1\nu}q_{\mu}-p_{\mu}l_{1\nu}-p_{\nu}l_{1\mu} -2 l_{1\mu}l_{1\nu}-l_{1\mu}l_{2\nu}-l_{1\nu}l_{2\mu}\right)\Bigr)\Bigr),
\end{eqnarray}
\begin{eqnarray}
\Gamma_{k\, \phi\, h_{\mu \nu}^{TT}}&=& -\frac{Z_{\phi}(k)}{2}\phi(q-p) \left(2 p_{\mu}p_{\nu}-q_{\mu}p_{\nu} -q_{\nu}p_{\mu}+q_{\mu}q_{\nu}\right)\nonumber\\
&+& \int_{l_1, l_2}\phi(l_1)\phi(l_2)\phi(q-p-l_1-l_2)\Bigl(-2 \bar{\rho}_b(k) \left(2 l_{1\mu}l_{1\nu}l_{2}^2+p_{\mu}p_{\nu}l_2^2+l_{2\mu}l_{2\nu}p^2 \right)\nonumber\\
&{}&
-2 \bar{\rho}_a(k)\left(\left(p_{\mu}l_{1\, \nu}+p_{\nu}l_{1\, \mu}\right)(l_2 \cdot q - l_2\cdot p -l_1 \cdot l_2 -l_2^2)+\left(l_{1\, \mu}l_{2\, \nu}+l_{1\, \nu}l_{2\, \mu} \right)\left( p \cdot q - p^2-p \cdot l_1 -p \cdot l_2\right) \right)\nonumber\\
&{}&-\frac{\bar{\rho}_c}{2}  \Bigl(l_{2\mu}l_{2\nu}\left(2 l_1 \cdot p+2l_1 \cdot q-2l_1^2-2 l_1 \cdot l_2 \right)+p^2 \left(l_{1\mu}l_{2\nu}+l_{1\nu}l_{2\mu} \right)+2 p_{\mu}p_{\nu}l_1 \cdot l_2 \nonumber\\
&{}&\quad \quad + l_2^2 \left(l_{1\mu}p_{\nu}+l_{1\nu}p_{\mu}+q_{\mu}l_{1\nu}+q_{\nu}l_{1\mu} -2 l_{1\mu}l_{1\nu}-l_{1\mu}l_{2\nu}-l_{1\nu}l_{2\mu}\right)\Bigr)
\Bigr),
\end{eqnarray}
\begin{eqnarray}
&{}&\Gamma_{k\, h_{\mu \nu}^{TT}\, h_{\kappa \lambda}^{TT}}^{(2)}(p,q)\nonumber\\
&=&\frac{Z_{\phi}(k)}{8} \int_{l_1}\phi(l_1)\,\phi(q-p-l_1)\Bigl(l_1^{\gamma}(-p_{\gamma}+q_{\gamma}-l_{1\, \gamma})\left(\delta_{\mu \kappa}\delta_{\nu \lambda}+\delta_{\mu \lambda}\delta_{\nu \kappa} \right)  \nonumber\\
&{}& \phantom{xxxx}+ \Bigl( \left(l_{1\, \mu}\delta_{\nu \lambda}+l_{1\, \nu}\delta_{\mu \lambda} \right)(p_{\kappa}+l_{1\, \kappa})
+\left(l_{1\, \mu}\delta_{\nu \kappa}+l_{1\, \nu}\delta_{\mu \kappa} \right)(p_{\lambda}+l_{1\, \lambda})\nonumber\\
&{}&\phantom{xxxxxx}
+\left(l_{1\, \kappa}\delta_{\lambda \mu}+ l_{1\, \lambda}\delta_{\kappa \mu} \right)\left (-q_{\nu}+l_{1\,\nu} \right)
+\left(l_{1\, \kappa}\delta_{\lambda \nu}+ l_{1\, \lambda}\delta_{\kappa \nu} \right)\left(-q_{\mu}+l_{1\,\mu} \right)
\Bigr)
\Bigr)\nonumber\\
&+& \int_{l_1,l_2,l_3} \phi(l_1)\,\phi(l_2)\,\phi(l_3)\,\phi(q-p-l_1-l_2-l_3)\Bigl(\frac{\bar{\rho}_a(k)}{2}\Bigl[ \frac{-1}{2}\left(\delta_{\mu \kappa}\delta_{\nu \lambda}+ \delta_{\mu \lambda}\delta_{\nu \kappa} \right)l_1 \cdot l_2 (l_3 \cdot q - l_3 \cdot p -l_1\cdot l_3 -l_2 \cdot l_3 -l_3^2)\nonumber\\
&{}& \phantom{\frac{\bar{\rho}_{\phi}(k)}{2}\int_{l_1,l_2,l_3} }+ \Bigl( l_{1\, \mu}l_{2\, \kappa}\delta_{\nu \lambda}+l_{1\, \mu}l_{2\, \lambda}\delta_{\nu \kappa}+l_{1\, \nu}l_{2\, \kappa}\delta_{\mu \lambda}+l_{1\, \nu}l_{2\, \lambda}\delta_{\mu \kappa}+l_{1\, \kappa}l_{2\, \nu}\delta_{\mu \lambda}+l_{1\, \kappa}l_{2\, \mu}\delta_{\nu \lambda}+ l_{1\,\lambda}l_{2\, \mu}\delta_{\nu \kappa}+ l_{1\, \lambda}l_{2\, \nu}\delta_{\mu \kappa} \Bigr)\cdot\nonumber\\
&{}& \phantom{\frac{\bar{\rho}_{\phi}(k)}{2}\int_{l_1,l_2,l_3}} \cdot (l_3 \cdot q - l_3 \cdot p -l_1 \cdot l_3 -l_2 \cdot l_3 -l_3^2)\nonumber\\
&{}& \phantom{\frac{\bar{\rho}_{\phi}(k)}{2}\int_{l_1,l_2,l_3} }+ \frac{1}{2}\Bigl(\left(l_{1\, \mu}l_{2\, \nu}+l_{1\, \nu}l_{2\, \mu} \right)\left(l_{3\, \kappa}\left(-p_{\lambda}-l_{1\, \lambda}-l_{2\, \lambda}-l_{3\, \lambda} \right)+l_{3\, \lambda}\left( -p_{\kappa}-l_{1\, \kappa}-l_{2\, \kappa}-l_{3\, \kappa}\right) \right)\nonumber\\
&{}&\phantom{\frac{\bar{\rho}_{\phi}(k)}{2}\int_{l_1,l_2,l_3}}+\left(l_{1\, \kappa}l_{2\, \lambda}+ l_{1\, \lambda}l_{2\, \kappa} \right)\left(l_{3\, \mu}\left( q_{\nu}-l_{1\, \nu}-l_{2\, \nu}-l_{3\, \nu}\right) +l_{3\, \nu}\left( q_{\mu}-l_{1\, \mu}-l_{2\, \mu}-l_{3\, \mu}\right)\right) \Bigr)
\Bigr]\nonumber\\
&{}&\phantom{\frac{\bar{\rho}_{\phi}(k)}{2}}+ \frac{\bar{\rho}_b}{2} \Bigl[ \left(-l_1^2 l_2^2 +2 q \cdot l_1 l_2^2-2 p \cdot l_1 l_2^2\right)\frac{1}{2}\left(\delta_{\mu \kappa}\delta_{\nu \lambda}+ \delta_{\mu \lambda}\delta_{\nu \kappa} \right)\nonumber\\
&{}&\phantom{\frac{\bar{\rho}_{\phi}(k)}{2}\int_{l_1,l_2,l_3}}+l_2^2 \left(\left(p_{\lambda}l_{1\mu}-q_{\mu}l_{1\lambda}\right)\delta_{\nu \kappa}+\left(p_{\kappa}l_{1\mu}-q_{\mu}l_{1\kappa}\right)\delta_{\nu \lambda}+\left(p_{\lambda}l_{1\nu}-q_{\nu}l_{1\lambda}\right)\delta_{\mu \kappa}+\left(p_{\kappa}l_{1\nu}-q_{\nu}l_{1\kappa}\right)\delta_{\mu \lambda}  \right)\nonumber\\
&{}&\phantom{\frac{\bar{\rho}_{\phi}(k)}{2}\int_{l_1,l_2,l_3}} +2l_2^2 \left(l_{1\mu}l_{1\lambda}\delta_{\nu \kappa}+ l_{1\mu}l_{1\kappa}\delta_{\nu \lambda}+ l_{1\nu}l_{1\lambda}\delta_{\kappa \mu}+ l_{1\nu}l_{1\kappa}\delta_{\lambda \mu} \right) + 2 \left(l_{1\mu}l_{1\nu}l_{2\kappa}l_{2\lambda}+l_{1\kappa}l_{1\lambda}l_{2\mu}l_{2\nu} \right) \Bigr]\nonumber\\
&{}&\phantom{\frac{\bar{\rho}_{\phi}(k)}{2}}+ \frac{\bar{\rho}_c}{2} \Bigl[ \left( - l_1 \cdot l_2 l_3^2 + l_1 \cdot l_2 q \cdot l_3 - l_1 \cdot l_2 p \cdot l_3\right))\frac{1}{2}\left(\delta_{\mu \kappa}\delta_{\nu \lambda}+ \delta_{\mu \lambda}\delta_{\nu \kappa} \right)\nonumber\\
&{}&\phantom{\frac{\bar{\rho}_{\phi}(k)}{2}\int_{l_1,l_2,l_3}}+\frac{1}{2} l_3^2 \left( \left(l_{1\mu}l_{2\lambda}+l_{1\lambda}l_{2\mu}\right)\delta_{\nu \kappa}+ \left(l_{1\mu}l_{2\kappa}+l_{1\kappa}l_{2\mu} \right)\delta_{\lambda \nu}+ \left( l_{1\nu}l_{2 \lambda}+ l_{1\lambda}l_{2\nu}\right)\delta_{\mu \kappa}+ \left(l_{1}\nu l_{2\kappa}+l_{1\kappa}l_{2\nu}\right)\delta_{\lambda \mu}\right)\nonumber\\
&{}&\phantom{\frac{\bar{\rho}_{\phi}(k)}{2}\int_{l_1,l_2,l_3}}+ l_1 \cdot l_2 \left( l_{3\mu}l_{3\lambda}\delta_{\nu \kappa}+ l_{3\mu}l_{3\kappa}\delta_{\nu \lambda}+ l_{3\nu}l_{3\lambda}\delta_{\mu \kappa}+ l_{3\nu}l_{3\kappa}\delta_{\mu \lambda}\right)\nonumber\\
&{}&\phantom{\frac{\bar{\rho}_{\phi}(k)}{2}\int_{l_1,l_2,l_3}}+\frac{1}{2} l_1 \cdot l_2 \left(\left(p_{\lambda}l_{3\mu}-q_{\mu}l_{3\lambda} \right)\delta_{\nu \kappa}+ \left(p_{\lambda}l_{3\nu}-q_{\nu}l_{3\lambda} \right)\delta_{\mu \kappa} +\left(p_{\kappa}l_{3 \mu}-q_{\mu}l_{3\kappa} \right)\delta_{\lambda \nu}+ \left(p_{\kappa}l_{3\nu}-q_{\nu}l_{3\kappa} \right)\delta_{\mu \lambda}\right)
\nonumber\\
&{}&\phantom{\frac{\bar{\rho}_{\phi}(k)}{2}\int_{l_1,l_2,l_3}}+ \left(l_{1\mu}l_{2\nu} +l_{1\nu}l_{2\mu}\right)l_{3\kappa}l_{3\lambda}+ \left( l_{1\kappa}l_{2\lambda}+l_{1\lambda}l_{2\kappa}\right)l_{3\mu}l_{3\nu}
\Bigr]\nonumber\\
&{}& \phantom{\frac{\bar{\rho}_{\phi}(k)}{2}} - \frac{\bar{\lambda}_{\phi}}{4} \left(\delta_{\mu \kappa}\delta_{\nu \lambda}+ \delta_{\mu \lambda}\delta_{\nu \kappa} \right)
\Bigr).
\end{eqnarray}
Finally

\begin{eqnarray}
\Gamma_{k\, \phi \phi}^{(2)}(p,q)
&=& \int_{l_1} \phi(l_1)\,\phi(q-p-l_1) \Bigl( \bar{\rho}_a(k)\left(
-8 p \cdot q\, l_1 \cdot q +8 p\cdot q\, l_1 \cdot p + 4 p \cdot q\, l_1^2+ 8 p \cdot l_1 \,q \cdot l_1 -4 p \cdot l_1\, q^2 +4 l_1 \cdot q \,p^2 \right)\nonumber\\
&{}&\bar{\rho}_b(k)\left(2 l_1^2 \left(q^2+p^2+l_1^2-2 q \cdot p -2 q \cdot l_1+ 2 p \cdot l_1\right) +4 q^2 l_1^2+4 p^2 l_1^2+2 p^2 q^2\right)\nonumber\\
&{}&\bar{\rho}_c(k) \left(-3 p^2 l_1^2-2 p\cdot l_1 l_1^2+p \cdot l_1 q^2-3 q^2 l_1^2+2 q \cdot l_1 l_1^2 -q \cdot l_1 p^2+2p\cdot q l_1^2+q \cdot l_1 q^2 -p \cdot l_1 p^2 \right)\nonumber\\
&{}& +12 \bar{\lambda}_{\phi}
\Bigr).
\end{eqnarray}

For the graviton propagators, see \cite{Eichhorn:2010tb}. The scalar propagator is given by $\mathcal{P}_{\phi \phi}(p,q)= Z_{\phi}(k) p^2 \delta^{4}(p-q)$.
\end{widetext}

\section{Matter contribution to gravitational $\beta$ functions}\label{matterbetacontr}
To obtain the matter contribution to the gravitational $\beta$ functions, we evaluate the flow equation on a spherical background, where $\bar{D}^2$ is the Laplacian acting on scalar fields, and the expansion of the heat-kernel trace is well-known:
\begin{equation}
e^{-s \left( -\bar{D}^2\right)}=\left( \frac{1}{4 \pi s}\right)^2 \int d^4x \sqrt{g}\, \left(1+ \frac{1}{6}s\, R+ ...  \right).\label{heatK}
\end{equation}
The first two terms in the expansion suffice to project $\partial_t \Gamma_k$ onto $\beta_{g}$ and $\beta_{\lambda}$. 

To use this, we rewrite $\partial_t \Gamma_{k\, \phi}$, using inverse Laplace transformations for shape functions as obtained in \cite{Gies:2002af}, where the subscript $\phi$ implies that we only consider scalar fluctuation fields, as follows\newline\\
$\phantom{x}$\\
$\phantom{x}$\\
\begin{eqnarray}
\partial_t \Gamma_{k\, \phi} &=& \frac{1}{2} {\rm Tr} \frac{\partial_t R_k}{\Gamma_k^{(2)}+ R_k} \nonumber\\
&=&\frac{1}{2} {\rm Tr} \left(\frac{\partial_t \Gamma_k^{(2)}}{\Gamma_k} \frac{r(y)}{1+r(y)} \right)+ \frac{1}{2}{\rm Tr} \frac{r'(y) (-2 y)}{1+r(y)}\nonumber\\
&=& - \frac{\eta_{\phi}}{2} \int_0^{\infty} ds\, \delta(s-1) \,{\rm Tr}\, e^{-s\frac{-\bar{D}^2}{k^2}} \nonumber\\
&{}&-\int_0^{\infty}ds\, \sum_{m=1}^{\infty} \delta(s-m)\frac{\rm d}{{\rm d}s}\, {\rm Tr} \,e^{-s\frac{-\bar{D}^2}{k^2}} ,
\end{eqnarray}
where we explicitly used $r(y)= \left( e^{y}-1\right)^{-1}$.

In the next step, we use the heat-kernel expansion  \Eqref{heatK} to obtain the matter contributions to the $\beta$ functions
\begin{eqnarray}
\partial_t g\vert_{\rm matter}&=& 16 \pi \, g^2 \left(-\frac{1}{12 \,(4 \pi)^2}\eta_{\phi} +\frac{1}{16 \cdot 36}\right)\nonumber\\
\partial_t \lambda\vert_{\rm matter}&=&  g \frac{\zeta(3)}{ \pi}+  16 \pi\, g \, \lambda\,\left( -\frac{1}{12\cdot 16 \pi^2}\eta_{\phi}+ \frac{1}{16 \cdot 36}\right)\nonumber\\
&{}&- \frac{\eta_{\phi}\, g}{4 \pi}.
\end{eqnarray}

\section{Scale-derivative of the regulator}\label{partialtreg}
After projecting the right-hand side of the Wetterich equation on the appropriate power of external momenta, it contains up to the fourth derivative of the regulator shape function $r(y)$. Extending \cite{Eichhorn:2010tb}, we observe that for a function $f$ depending on the regulator shape function $r_i$ (here $i$ is an index that labels the different fields) and its derivatives the following holds:
\begin{widetext}
\begin{eqnarray}
&{}&\tilde{\partial}_t f(r_i,r_i^{(1)},...,r_i^{(4)})(p)\nonumber\\
&=& \int_{p'} \partial_t R_i(p') \frac{\delta}{\delta R_i(p')} f(r_i,r_i^{(1)},...,r_i^{(4)})(p)\nonumber\\
&=&\int_{p'} \frac{\partial_t R_i(p')}{Z_i\left(p'^2-\lambda_i \right)}\left( f^{(1,0,0,0,0)}(r_i,...,r_i^{(4)})(p) \delta_{ij}\delta_{p,p'}+...+f^{(0,0,0,0,1)}(r_i,...,r_i^{(4)})(p) \delta_{ij}\partial_{\tilde{y}}^4\delta_{p,p'}\right)\nonumber\\
&=&f^{(1,0,0,0,0)}(r_i,r_i^{(1)},...,r_i^{(4)})(p)  \frac{\partial_t R_i(p)}{Z_i\left(p^2-\lambda_i \right)}+...+f^{(0,0,0,0,1)}(r_i,r_i^{(1)},...,r_i^{(4)})(p)  \partial_{\tilde{y}}^4 \frac{\partial_t R_i(p)}{Z_i\left(p^2-\lambda_i \right)},
\end{eqnarray}
wherein $\tilde{y}= \frac{p^2- \lambda_i}{k^2}$ is the argument of the regulator shape function. Herein $\lambda_i=0$ for the scalar matter shape function. Rewriting $\partial_{\tilde{y}}= \frac{k^2}{2p^2}p_{\mu}\frac{\partial}{\partial p_{\mu}}$ allows to deduce the form of $\partial_{\tilde{y}}^n \frac{\partial_t R)i(p)}{Z_i\left(p^2-\lambda_i \right)}$ straightforwardly.
\end{widetext}

\section{$\beta$ functions for unspecified shape function}\label{matterbetaunspecifiedr}

Here we present the $\beta$ functions for the dimensionfull (unrenormalised) quantities for a regulator of the form $\Gamma_k^{(2)}\Big|_{\phi=0}r(\frac{p^2}{k^2})$. All numerical results in the main text are obtained using an exponential shape function.
\begin{widetext}
\begin{eqnarray}
\beta_{\bar{\rho}_a}&=&\frac{1}{2}\int \frac{d^4p}{(2 \pi)^4} \Bigl[\left( \frac{10}{3} \brhoa + \frac{10}{9} \brhob-\frac{5}{9}\brhoc\right)\patt\mathcal{G}_{TT} -\frac{1}{2} \frac{575}{1728} Z_{\phi}^2\patt \mathcal{G}_{TT}^2\nonumber\\
&{}& -\frac{1}{2}\Bigl[ \left(40 (p^2)^2 \brhoa^2- 80 (p^2)^2 \brhoa \brhob+ 16 (p^2)^2 \brhob^2+ 48 \blam \brhoc- \frac{40}{3}(p^2)^2 \brhoa \brhoc+ 32 (p^2)^2 \brhob \brhoc+ 2 (p^2)^2 \brhoc^2 \right)\patt\mathcal{G}_{\phi}^2\nonumber\\
&{}&\, \, \, \,\,\,+
\left( 480 p^2\blam \brhoa -192 p^2 \blam \brhob+ \frac{400}{3} (p^2)^3 \brhoa \brhob - 80 (p^2)^3 \brhob^2- 144 \blam p^2 \brhoc- 40 (p^2)^3 \brhob \brhoc\right) \patt \mathcal{R}^{(1)}\mathcal{G}_{\phi}^3\nonumber\\
&{}& \, \, \, \,\,\,+
\left( -288 \blam^2 + 160 (p^2)^2 \blam \brhoa - 384 \blam (p^2)^2 \brhob -48 (p^2)^2 \blam \brhoc+\frac{80}{3}(p^2)^4\brhoa \brhob-72 (p^2)^4 \brhob^2- 8 (p^2)^4 \brhob \brhoc\right) \patt \mathcal{R}^{(2)}\mathcal{G}_{\phi}^3\nonumber\\
&{}&\, \, \, \,\,\,+
\left( -288 p^2 \blam^2-160 (p^2)^3 \blam \brhob -\frac{56}{3}(p^2)^5 \brhob^2\right)\patt \mathcal{R}^{(3)} \mathcal{G}_{\phi}^3 +
\left(-48 \blam^2 (p^2)^2 - 16 \blam (p^2)^4 \brhob - \frac{4}{3} (p^2)^6 \brhob^2\right) \patt \mathcal{R}^{(4)}\mathcal{G}_{\phi}^3\nonumber\\
&{}&\, \, \, \,\,\,+ \left(\!576 \blam^2-\!320 (p^2)^2 \blam \brhoa +\!768 \blam (p^2)^2 \brhob -\! \frac{160}{3} (p^2)^4 \brhoa \brhob +\! 144 (p^2)^4 \brhob^2 +\!96 \blam (p^2)^2 \brhoc + \!16 (p^2)^4 \brhob \brhoc \!\right)\patt (\mathcal{R}^{(1)} )^2\mathcal{G}_{\phi}^4\nonumber\\
&{}&\, \, \, \,\,\,+\left( 288 \blam^2 (p^2)^2 +96 \blam(p^2)^4 \brhob + 8 (p^2)^6 \brhob^2\right)\patt \left(\mathcal{R}^{(2)} \right)^2\mathcal{G}_{\phi}^4
+\left(1728 \blam^2 p^2 +960 (p^2)^3  \blam \brhob + 112 (p^2)^5 \brhob^2\right)\patt \mathcal{R}^{(1)} \mathcal{R}^{(2)}\mathcal{G}_{\phi}^4\nonumber\\
&{}&\, \, \, \,\,\, +\left( 384 \blam^2 (p^2)^2 + 128 \blam (p^2)^4 \brhob + \frac{32}{3} (p^2)^6 \brhob^2\right)\patt \mathcal{R}^{(1)} \mathcal{R}^{(3)}\mathcal{G}_{\phi}^4\nonumber\\
&{}&\, \, \, \,\,\,+\left( -1728 \blam^2 p^2 -960 \blam (p^2)^3 \brhob - 112 (p^2)^5 \brhob^2\right) \patt \left(\mathcal{R}^{(1)}\right)^3 \mathcal{G}_{\phi}^5\nonumber\\
&{}& \, \, \, \,\,\,+  \left(-1728 \blam^2 (p^2)^2 -576 \blam (p^2)^4 \brhob - 48 (p^2)^6 \brhob^2 \right) \patt \left(\mathcal{R}^{(1)}\right)^2 \mathcal{R}^{(2)} \mathcal{G}_{\phi}^5\nonumber\\
 &{}&\, \, \, \,\,\,+\left(1152 \blam^2 (p^2)^2+384 \blam (p^2)^4 \brhob +32 (p^2)^6 \brhob^2 \right)\patt \left(\mathcal{R}^{(1)}\right)^4\mathcal{G}_{\phi}^6\Bigr]\nonumber\\
&{}&- \frac{1}{2} Z{\phi}\left( \frac{20}{9} \brhob p^2 - \frac{5}{9} \brhoc p^2\right) \patt \mathcal{G}_{TT} \mathcal{G}_{\phi \phi}
- \frac{1}{2} Z{\phi}\left( -\frac{1}{96} \brhoc p^4 \right) \patt \mathcal{G}_{hh} \mathcal{R}^{(1)}\mathcal{G}_{\phi \phi}^2\nonumber\\
&{}& +\frac{1}{3}Z_{\phi}^2\Bigl[ \left( \frac{3}{8} \brhoa(p^2)^2 + \frac{1}{8}\brhob (p^2)^2- \frac{1}{16} \brhoc (p^2)^2\right)\patt \mathcal{G}_{hh} \mathcal{G}_{\phi \phi}^2
+ \left( -\frac{1}{4} \brhob (p^2)^3\right) \patt \mathcal{G}_{hh} \mathcal{G}_{\phi \phi}^3 \mathcal{R}^{(1)}\nonumber\\
&{}& \, \, \, \,\,\,+ \left( \frac{3}{4} \blam (p^2)^2 + \frac{1}{8} \brhob (p^2)^4\right) \patt \mathcal{G}_{hh} \mathcal{G}_{\phi \phi}^4 \left(\mathcal{R}^{(1)} \right)^2\Bigr]
+\frac{1}{3} Z_{\phi}^2 \left( 20 \blam + \frac{10}{3} \brhob (p^2)^2\right) \patt \mathcal{G}_{TT} \mathcal{G}_{\phi \phi}^2\nonumber\\
&{}& -\frac{1}{4} \frac{1}{1024}Z_{\phi}^4 \patt \mathcal{G}_{\phi\phi}^2 \mathcal{G}_{hh}^2
\Bigr].
\end{eqnarray}

\begin{eqnarray}
\beta_{\bar{\rho}_b}&=&\frac{1}{2}\int \frac{d^4p}{(2 \pi)^4} \Bigl[ \frac{10}{3} \bar{\rho}_b \patt \mathcal{G}_{TT}-\frac{1}{2} \frac{85}{576} Z_{\phi}^2\patt \mathcal{G}_{TT}^2
\nonumber\\
&{}& -\frac{1}{2}\Bigl[ \left(240 \blam \brhob +216 (p^2)^2 \brhob^2 - 72  (p^2)^2 \brhob \brhoc+10 \brhoc^2 (p^2)^2 \right)\patt\mathcal{G}_{\phi}^2\nonumber\\
&{}& \, \, \, \,\,\,+
\left(  -768 \blam \brhob p^2-240 \brhob^2 (p^2)^3+144 \blam \brhoc p^2+40 \brhob \brhoc (p^2)^3\right) \patt \mathcal{R}^{(1)}\mathcal{G}_{\phi}^3\nonumber\\
&{}& \, \, \, \,\,\,+
\left( -288 \blam^2-576 (p^2)^2 \blam \brhob -104 \brhob^2 (p^2)^4 +48 \blam \brhoc (p^2)^2+8 \brhob \brhoc (p^2)^4\right) \patt \mathcal{R}^{(2)}\mathcal{G}_{\phi}^3\nonumber\\
&{}& \, \, \, \,\,\,+
\left( -288 \blam^2  p^2-160 \blam \brhob (p^2)^3 - \frac{56}{3} \brhob^2 (p^2)^5\right) \patt \mathcal{R}^{(3)}\mathcal{G}_{\phi}^3+
\left( -48 \blam^2 (p^2)^2 -16 \blam \brhob (p^2)^4 - \frac{4}{3} \brhob^2 (p^2)^6\right) \patt \mathcal{R}^{(4)}\mathcal{G}_{\phi}^3\nonumber\\
&{}&\, \, \, \,\,\,+ \left( 576 \blam^2+1152 \blam \brhob (p^2)^2+208 \brhob^2 (p^2)^4-96 \blam \brhoc (p^2)^2-16 \brhob \brhoc (p^2)^4\right)\patt \left(\mathcal{R}^{(1)} \right)^2\mathcal{G}_{\phi}^4\nonumber\\
&{}&\, \, \, \,\,\,+\left(288 \blam^2 (p^2)^2+96 \blam \brhob (p^2)^4+8 \brhob^2 (p^2)^6\right)\patt \left(\mathcal{R}^{(2)} \right)^2\mathcal{G}_{\phi}^4\nonumber\\
&{}&\, \, \, \,\,\,
+\left( 1728 \blam^2 p^2+960 \blam \brhob (p^2)^3+112 \brhob^2 (p^2)^5\right)\patt \mathcal{R}^{(1)} \mathcal{R}^{(2)}\mathcal{G}_{\phi}^4\nonumber\\
&{}&\, \, \, \,\,\, +\left(384 \blam^2 (p^2)^2+128 \blam \brhob (p^2)^4+\frac{32}{3} \brhob^2 (p^2)^6\right)\patt \mathcal{R}^{(1)} \mathcal{R}^{(3)}\mathcal{G}_{\phi}^4\nonumber\\
&{}&\, \, \, \,\,\,+\left( -1728 \blam^2 p^2-960 \blam \brhob (p^2)^3-112 \brhob^2 (p^2)^5\right) \patt \left(\mathcal{R}^{(1)}\right)^3 \mathcal{G}_{\phi}^5\nonumber\\
&{}& \, \, \, \,\,\,+  \left( -1728 \blam^2 (p^2)^2-576 \blam \brhob (p^2)^4-48 \brhob^2 (p^2)^6\right) \patt \left(\mathcal{R}^{(1)}\right)^2 \mathcal{R}^{(2)} \mathcal{G}_{\phi}^5\nonumber\\
&{}&\, \, \, \,\,\,+\left(1152 \blam^2 (p^2)^2 + 384 \blam \brhob (p^2)^4 + 32 \brhob^2 (p^2)^6\right)\patt \left(\mathcal{R}^{(1)}\right)^4\mathcal{G}_{\phi}^6\Bigr]\nonumber\\
&{}&- \frac{1}{2} Z{\phi}\left( \frac{5}{3} \brhob p^2 \right) \patt \mathcal{G}_{TT} \mathcal{G}_{\phi \phi}\nonumber\\
&{}&- \frac{1}{2}Z{\phi} \Bigl[\left( -\frac{1}{16} \brhob p^2\right)\patt \mathcal{G}_{hh} \mathcal{G}_{\phi \phi}
+\left( -3 \blam- \frac{3}{32} \brhob (p^2)^2 \right) \patt \mathcal{G}_{hh} \mathcal{R}^{(1)}\mathcal{G}_{\phi \phi}^2
+\left( -3 \blam p^2- \frac{1}{32} \brhob (p^2)^3 \right) \patt \mathcal{G}_{hh} \mathcal{R}^{(2)}\mathcal{G}_{\phi \phi}^2\nonumber\\
&{}&\, \, \, \,\,\,
+\left(  -\frac{1}{2}\blam (p^2)^2\right) \patt \mathcal{G}_{hh} \mathcal{R}^{(3)}\mathcal{G}_{\phi \phi}^2
+\left(+6 \blam p^2  +\frac{1}{16}\brhob\right) \patt \mathcal{G}_{hh} \left(\mathcal{R}^{(1)}\right)^2\mathcal{G}_{\phi \phi}^3
+\left(  3 \blam (p^2)^2\right) \patt \mathcal{G}_{hh} \mathcal{R}^{(1)}\mathcal{R}^{(2)}\mathcal{G}_{\phi \phi}^3\nonumber\\
&{}&\, \, \, \,\,\,
+\left(-3 \blam (p^2)^2\right) \patt \mathcal{G}_{hh} \left(\mathcal{R}^{(1)}\right)^3\mathcal{G}_{\phi \phi}^4
\Bigr]\nonumber\\
&{}& +\frac{1}{3}Z_{\phi}^2\Bigl[ \left( \frac{9}{4} \blam+ \frac{3}{2} \brhob (p^2)^2\right)\patt \mathcal{G}_{hh} \mathcal{G}_{\phi \phi}^2
+ \left( - \frac{45}{8}\blam p^2 - \frac{27}{16} \brhob (p^2)^3\right) \patt \mathcal{G}_{hh} \mathcal{G}_{\phi \phi}^3 \mathcal{R}^{(1)}\nonumber\\
&{}& \, \, \, \,\,\,+ \left( - \frac{9}{8} \blam (p^2)^2- \frac{3}{16} \brhob (p^2)^4\right) \patt \mathcal{G}_{hh} \mathcal{G}_{\phi \phi}^3 \mathcal{R}^{(2)}+ \left(\frac{27}{8} \blam (p^2)^2 + \frac{9}{16} \brhob (p^2)^4 \right) \patt \mathcal{G}_{hh} \mathcal{G}_{\phi \phi}^4 \left(\mathcal{R}^{(1)} \right)^2\Bigr]
\nonumber\\
&{}&\frac{1}{3} Z_{\phi}^2 \left( 15 \blam + \frac{5}{2} \brhob (p^2)^2\right) \patt \mathcal{G}_{TT} \mathcal{G}_{\phi \phi}^2
\Bigr].
\end{eqnarray}

\begin{eqnarray}
\beta_{\bar{\rho}_c}&=&\frac{1}{2}\int \frac{d^4p}{(2 \pi)^4} \Bigl[ \left( \frac{10}{3} \bar{\rho}_b+\frac{5}{3} \bar{\rho}_c\right) \patt \mathcal{G}_{TT}-\frac{1}{2} \frac{55}{96} Z_{\phi}^2\patt \mathcal{G}_{TT}^2
\nonumber\\
&{}& -\frac{1}{2}\Bigl[ \left(192 \blam \brhob - 144 \brhoa \brhob (p^2)^2 +160 \brhob^2  (p^2)^2+144 \blam \brhoc +40 \brhoa \brhoc  (p^2)^2+40 \brhob \brhoc  (p^2)^2-4 \brhoc^2  (p^2)^2\right)\patt\mathcal{G}_{\phi}^2\nonumber\\
&{}&\, \, \, \,\,\,+
\left( 288 \blam \brhoa p^2 -960 \blam \brhob  p^2 +80 \brhoa \brhob  (p^2)^3 -320 \brhob^2  (p^2)^3\right) \patt \mathcal{R}^{(1)}\mathcal{G}_{\phi}^3\nonumber\\
&{}& \, \, \, \,\,\,+
\left(-176 \brhob^2  (p^2)^4 +16 \brhoa \brhob  (p^2)^4 -960 \blam \brhob  (p^2)^2 -576 \blam^2  +96 \blam \brhoa (p^2)^2\right) \patt \mathcal{R}^{(2)}\mathcal{G}_{\phi}^3\nonumber\\
&{}& \, \, \, \,\,\,+
\left(-576 \blam^2 p^2 -320 \blam \brhob  (p^2)^3 -\frac{112}{3} \brhob^2  (p^2)^5\right) \patt \mathcal{R}^{(3)}\mathcal{G}_{\phi}^3\nonumber\\
&{}&\, \, \, \,\,\,+
\left( - \frac{8}{3} \brhob^2  (p^2)^6-96 \blam^2 (p^2)^2 -32 \blam \brhob (p^2)^4\right) \patt \mathcal{R}^{(4)}\mathcal{G}_{\phi}^3\nonumber\\
&{}&\, \, \, \,\,\,+ \left( 1152 \blam^2 -192 \blam \brhoa (p^2)^2+1920 \blam \brhob (p^2)^2 -32 \brhoa \brhob (p^2)^4 +352 \brhob^2 (p^2)^4\right)\patt \left(\mathcal{R}^{(1)} \right)^2\mathcal{G}_{\phi}^4\nonumber\\
&{}&\, \, \, \,\,\,+\left(576 \blam^2 (p^2)^2+192 \blam \brhob (p^2)^4 +16 \brhob^2 (p^2)^6\right)\patt \left(\mathcal{R}^{(2)} \right)^2\mathcal{G}_{\phi}^4\nonumber\\
&{}&\, \, \, \,\,\,
+\left(224 \brhob^2 (p^2)^5+1920 \blam \brhob (p^2)^3 +3456 \blam^2 p^2\right)\patt \mathcal{R}^{(1)} \mathcal{R}^{(2)}\mathcal{G}_{\phi}^4\nonumber\\
&{}&\, \, \, \,\,\, +\left(+\frac{64}{3} \brhob^2 (p^2)^6+256 \blam \brhob (p^2)^4+768 \blam^2 (p^2)^2\right)\patt \mathcal{R}^{(1)} \mathcal{R}^{(3)}\mathcal{G}_{\phi}^4\nonumber\\
&{}&\, \, \, \,\,\,+\left( -3456 \blam^2 p^2-224 \brhob^2 (p^2)^5-1920 \blam \brhob (p^2)^3\right) \patt \left(\mathcal{R}^{(1)}\right)^3 \mathcal{G}_{\phi}^5\nonumber\\
&{}& \, \, \, \,\,\,+  \left(-96 \brhob^2 (p^2)^6-3456 \blam^2 (p^2)^2 -1152 \blam \brhob (p^2)^4\right) \patt \left(\mathcal{R}^{(1)}\right)^2 \mathcal{R}^{(2)} \mathcal{G}_{\phi}^5\nonumber\\
 &{}&\, \, \, \,\,\,+\left(2304 \blam^2 (p^2)^2+768 \blam \brhob (p^2)^4+64 \brhob^2 (p^2)^6\right)\patt \left(\mathcal{R}^{(1)}\right)^4\mathcal{G}_{\phi}^6\Bigr]\nonumber\\
&{}&- \frac{1}{2}Z{\phi} \left( \frac{20}{3} \brhob p^2 - \frac{5}{6} \brhoc p^2\right) \patt \mathcal{G}_{TT} \mathcal{G}_{\phi \phi}\nonumber\\
&{}&- \frac{1}{2} Z{\phi}\Bigl[\left( \frac{1}{8} \brhob p^2- \frac{3}{32} \brhoc p^2\right)\patt \mathcal{G}_{hh} \mathcal{G}_{\phi \phi}
+\left( -6 \blam- \frac{5}{16}\brhob (p^2)^2 +\frac{1}{32} \brhoc (p^2)^2\right) \patt \mathcal{G}_{hh} \mathcal{R}^{(1)}\mathcal{G}_{\phi \phi}^2\nonumber\\
&{}&\, \, \, \,\,\,
+\left(  -6 \blam p^2- \frac{1}{16} \brhob (p^2)^3\right) \patt \mathcal{G}_{hh} \mathcal{R}^{(2)}\mathcal{G}_{\phi \phi}^2
- \blam (p^2)^2 \patt \mathcal{G}_{hh} \mathcal{R}^{(3)}\mathcal{G}_{\phi \phi}^2
+\left(12 \blam p^2+ \frac{1}{8} \brhob (p^2)^3\right) \patt \mathcal{G}_{hh} \left(\mathcal{R}^{(1)}\right)^2\mathcal{G}_{\phi \phi}^3\nonumber\\
&{}&\, \, \, \,\,\,
+\left( 6 \blam (p^2)^2 \right) \patt \mathcal{G}_{hh} \mathcal{R}^{(1)}\mathcal{R}^{(2)}\mathcal{G}_{\phi \phi}^3
+\left(-6 \blam (p^2)^2\right) \patt \mathcal{G}_{hh} \left(\mathcal{R}^{(1)}\right)^3\mathcal{G}_{\phi \phi}^4
\Bigr]\nonumber\\
&{}& +\frac{1}{3}Z_{\phi}^2\Bigl[ \left( \frac{9}{8}\brhob (p^2)^2+\frac{3}{16}\brhoc (p^2)^2\right)\patt \mathcal{G}_{hh} \mathcal{G}_{\phi \phi}^2
+ \left( - \frac{27}{4} \blam p^2 - \frac{21}{8} \brhob (p^2)^3\right) \patt \mathcal{G}_{hh} \mathcal{G}_{\phi \phi}^3 \mathcal{R}^{(1)}\nonumber\\
&{}& \, \, \, \,\,\,+ \left(- \frac{9}{4} \blam (p^2)^2- \frac{3}{8} \brhob (p^2)^4 \right) \patt \mathcal{G}_{hh} \mathcal{G}_{\phi \phi}^3 \mathcal{R}^{(2)}+ \left( \frac{27}{4} \blam (p^2)^2 + \frac{9}{8} \brhob (p^2)^4\right) \patt \mathcal{G}_{hh} \mathcal{G}_{\phi \phi}^4 \left(\mathcal{R}^{(1)} \right)^2\Bigr]\nonumber\\
&{}&\frac{1}{3} Z_{\phi}^2 \left( 60 \blam + 10 \brhob (p^2)^2\right) \patt \mathcal{G}_{TT} \mathcal{G}_{\phi \phi}^2
\Bigr].
\end{eqnarray}

\begin{eqnarray}
\beta_{\bar{\lambda}_{\phi}}&=&\frac{1}{2}\int \frac{d^4p}{(2 \pi)^4} \Bigl[ \left( -\frac{5}{2} \bar{\lambda}_{\phi}\right) \patt \mathcal{G}_{TT}+ \frac{1}{8} \bar{\lambda}_{\phi} \patt \mathcal{G}_{hh}
 -\frac{1}{2} \left(144 \blam^2 + 48 \blam \brhob (p^2)^2 + 4 \brhob^2 (p^2)^4\right)\patt\mathcal{G}_{\phi}^2
\Bigr].
\end{eqnarray}
\end{widetext}
Herein, $\mathcal{G}_{xx} = \frac{1}{\Gamma_{k\, xx}^{(2)}+R_k}$ and $\mathcal{R}^{(n)}= \frac{1}{(k^2)^{n-1}} \left(n r'(y)+ r''(y) y \right)$ for $n>1$ and $\mathcal{R}^{(1)}=1+r(y)+ r'(y)y$. Herein $r(y)$ is the regulator shape function with $y= \frac{p^2}{ k^2}$. This structure arises, since all diagrams where the external derivative with respect to the momentum hits the propagator, can be organised in a way that only derivatives of the scalar propagator, and not the graviton propagator, need to be taken.

\section{Scalar anomalous dimension}\label{anomdimapp}
Here we give our result for the anomalous dimension $\eta_{\phi}$ for a spectrally adjusted cutoff with exponential shape function:
\begin{widetext}
\begin{eqnarray}
\eta_{\phi}&=&\frac{1} {96 \pi  g \lambda  \left(e^{4 \lambda /3}
   \text{Ei}\left(\frac{4 \lambda }{3}\right)-\left(e^{4 \lambda /3}+1\right)
   \text{Ei}\left(\frac{8 \lambda }{3}\right)+\text{Ei}(4 \lambda )\right)+12 \pi  g
   e^{4 \lambda }-9 e^{8 \lambda /3} (4 \pi  (g+24 \pi )+9 (4 \rho_b-2
   \rho_c+3 \rho_a))} 4 e^{4 \lambda /3} \cdot \nonumber\\
&{}&\Bigl(8 \pi  g \left(\text{Ei}\left(\frac{8 \lambda
   }{3}\right) (\partial_t \lambda (3-12 \lambda )+\lambda  (3 \eta_N-32
   \lambda +6))+3 \text{Ei}(4 \lambda ) (\partial_t \lambda (4 \lambda -1)+\lambda 
   (-\eta_N+8 \lambda -2))+8 \lambda ^2 \text{Ei}\left(\frac{4 \lambda
   }{3}\right)\right)\nonumber\\
&{}&+4 e^{4 \lambda /3} \left(2 \pi  g \left(\text{Ei}\left(\frac{4
   \lambda }{3}\right)-\text{Ei}\left(\frac{8 \lambda }{3}\right)\right)
   (\partial_t \lambda (8 \lambda -3)+\lambda  (-3 \eta_N+16 \lambda -6))-3
   (\pi  g (4 \lambda +3)+9 (4 \rho_b-2 \rho_c+3 \rho_a))\right)\nonumber\\
&{}&-3 \pi
    g e^{8 \lambda /3} (4 \partial_t \lambda-3 \eta_N+6)-3 \pi  g \eta_N e^{4 \lambda }\Bigr)
\end{eqnarray}
\end{widetext}
\end{appendix}

\end{document}